\documentclass[12pt]{article}

\newcommand{\maintext}{1} 
\newcommand{\supplement}{1} 

\usepackage{amsmath}
\usepackage{amssymb}
\usepackage{amsthm}
\usepackage{array}
\usepackage{bm}
\usepackage{color}
\usepackage{comment}
\usepackage{fullpage}
\usepackage[margin=1in]{geometry}
\usepackage{graphicx}
\usepackage[hidelinks]{hyperref}
\usepackage{mathtools}
\usepackage{multirow}
\usepackage{natbib}
\usepackage{pifont} 
\usepackage{setspace}
\usepackage{subcaption}
\usepackage{textcomp} 
\usepackage{tikz}
\usepackage{url}
\usepackage{cleveref} 

\usetikzlibrary{shapes.geometric}

\newtheorem{corollary}{Corollary}
\newtheorem{proposition}{Proposition}
\newtheorem*{problem}{Problem Statement}

\newcommand{\tr}{\mathrm{tr}}
\newcommand{\defeq}{\vcentcolon=}
\newcommand{\eqdef}{=\vcentcolon}
\newcommand{\cmark}{\ding{51}}
\newcommand{\xmark}{\ding{55}}

\begin{document}

\title{A Statistical Approach to Surface Metrology for 3D-Printed Stainless Steel}
\author{Chris. J. Oates$^{1,2}$\footnote{Address for correspondence: Chris. J. Oates, School of Mathematics, Statistics and Physics, Herschel
Building, Newcastle University, Newcastle upon Tyne, NE1 7RU, UK. E-mail: \url{chris.oates@ncl.ac.uk}}, Wilfrid S. Kendall$^{3,2}$, Liam Fleming$^1$ \\ 
\small $^1$School of Mathematics, Statistics and Physics, Newcastle University, UK \\ 
\small $^2$Alan Turing Institute, UK \\ 
\small $^3$Department of Statistics, University of Warwick, UK}
\date{}
\maketitle

\if1\maintext
{

\begin{abstract}
Surface metrology is the area of engineering concerned with the study of geometric variation in surfaces.
This paper explores the potential for modern techniques from spatial statistics to act as generative models for geometric variation in 3D-printed stainless steel.
The complex macro-scale geometries of 3D-printed components pose a challenge that is not present in traditional surface metrology, as the training data and test data need not be defined on the same manifold.
Strikingly, a covariance function defined in terms of geodesic distance on one manifold can fail to satisfy positive-definiteness and thus fail to be a valid covariance function in the context of a different manifold; this hinders the use of standard techniques that aim to learn a covariance function from a training dataset.
On the other hand, the associated covariance differential operators are locally defined.
This paper proposes to perform inference for such differential operators, facilitating generalisation from the manifold of a training dataset to the manifold of a test dataset.
The approach is assessed in the context of model selection and explored in detail in the context of a finite element model for 3D-printed stainless steel.

\vspace{5pt}
\end{abstract}

\section{Introduction} \label{sec: intro}

The rate at which new manufacturing technologies are being developed and the range of components that can now be produced pose a challenge to formal standardisation, which has traditionally formed the basis for engineering design and assessment.
An important motivation for new technologies is to trade the quality of the component with the speed and cost at which it can be manufactured.
This trade-off results in the introduction of imperfections or defects, in a manner that is intrinsically unpredictable and could therefore be described as statistical.
As such, there is a need to develop principled and general statistical methods that can be readily adapted to new technologies, in order to obtain reliable models as the basis for formal standards in the engineering context.

The recent emergence of additive manufacture technologies for stainless steel promises to disrupt traditional approaches based on standardised components and to enable design of steel components of almost arbitrary size and complexity \citep{Buchanan2019}.
This \textit{directed energy deposition} technology arises from the combination of two separate technologies; traditional welding modalities and industrial robotic equipment \citep{ASTM}. 
In brief, material is produced in an additive manner by applying a layer of steel onto the surface of the material in the form of a continuous weld.
To ensure that this process is precisely controlled, the manufacture is performed by a robot on which the weld head is mounted.
The complexity of this process is such that the geometric (e.g. thickness and volume) and mechanical (e.g. stiffness and strength) properties of additively manufactured (henceforth ``3D-printed'') steel are not yet well-understood.
Indeed, conventional models for welding focus on the situation of a localised weld that is typically intended to bind two standard components together.
The material properties of elementary welds, as a function of the construction protocol and ambient environment, have been modelled in considerable detail \citep{Kou2003}.
In contrast, 3D-printed steel is a single global weld whose construction protocol is highly non-standard.
Furthermore, the limited precision of the equipment results in imperfections in the material geometry, which can range in severity from aesthetic roughness of the surface of the material through to macroscale defects, such as holes in the printed component.
These imperfections occur in a manner that is best considered as random, since they cannot be easily predicted or controlled.

The variation in the manufactured components precludes the efficient use of raw material, as large factors of safety are necessarily employed.
An important task is therefore to gain a refined understanding of the statistical nature of 3D-printed steel, enabling appropriate factors of safety to be provided for engineering design and assessment \citep{Buchanan2017}.
It is anticipated that geometric variation, as opposed to mechanical variation, is the dominant source of variation in the behaviour of a 3D-printed thin-walled steel component under loading.
The focus of this work is therefore to take a statistical approach to surface metrology, with the aim to develop a generative model for geometric variation in 3D-printed steel.
Specifically, the following problem is considered:
\begin{problem} 
Given the notional and actual geometries of a small number of 3D printed steel components, construct a generative statistical model for the actual geometry of a 3D-printed steel component whose notional geometry (only) is provided.
\end{problem}
\noindent For example, in \Cref{subsec: predict cylinder 1} of this paper a statistical model is used to generate realistic simulations of geometric variation that might be encountered on a 3D-printed steel cylinder, based only on a training dataset of notionally flat panels of 3D-printed steel; see \Cref{fig: explain}.

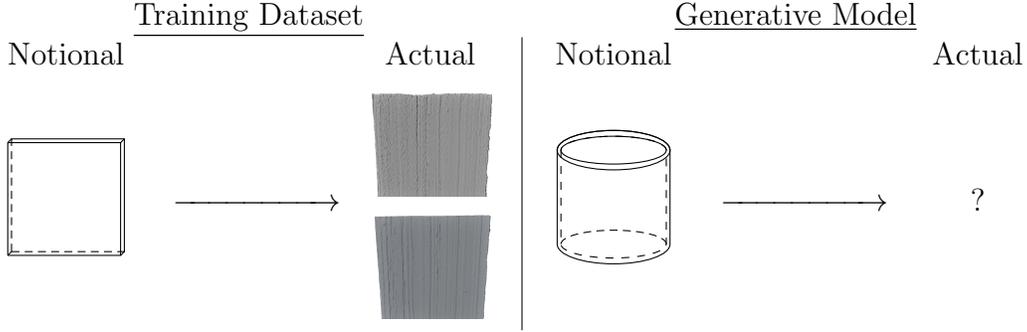
\begin{figure}[t!]
\centering

\newcolumntype{C}{ >{\centering\arraybackslash} m{2cm} }
\begin{tabular}{CCC|CCC}
\multicolumn{3}{c}{\underline{Training Dataset}} & \multicolumn{3}{c}{\underline{Generative Model}} \\
Notional & & Actual & Notional & & Actual \\
\begin{tikzpicture}
\draw [] (0,0) -- ++(1.5cm,0) -- ++ (0cm,1.5cm) -- ++ (-1.5cm,0) -- ++ (0cm,-1.5cm);
\draw [] (0.05,1.5) -- (0.05,1.55) -- (1.55,1.55) -- (1.55,0.05) -- (1.5,0.05);
\draw [densely dashed] (0.05,0.05) -- (0.05,1.5);
\draw [densely dashed] (0.05,0.05) -- (1.5,0.05);
\draw [densely dashed] (0,0) -- (0.05,0.05);
\draw [] (0,1.5) -- (0.05,1.55);
\draw [] (1.5,1.5) -- (1.55,1.55);
\draw [] (1.5,0) -- (1.55,0.05);
\end{tikzpicture}
&
$\xrightarrow{\hspace*{2cm}}$
&
\includegraphics[height = 0.07\textheight,clip,trim = 2cm 1.5cm 2cm 1cm]{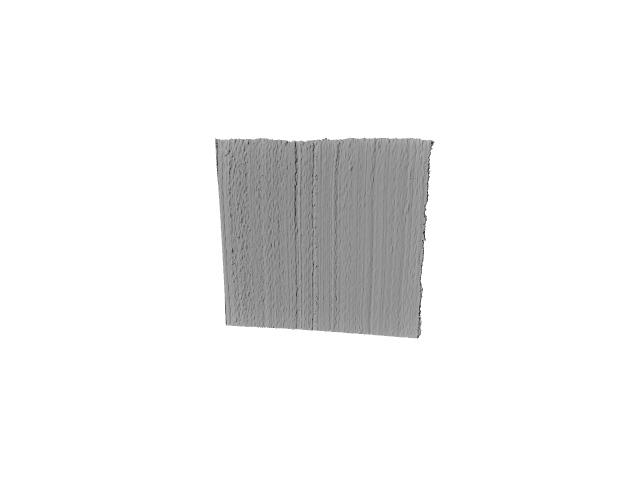} 
\includegraphics[height = 0.07\textheight,clip,trim = 2cm 1.5cm 2cm 1cm]{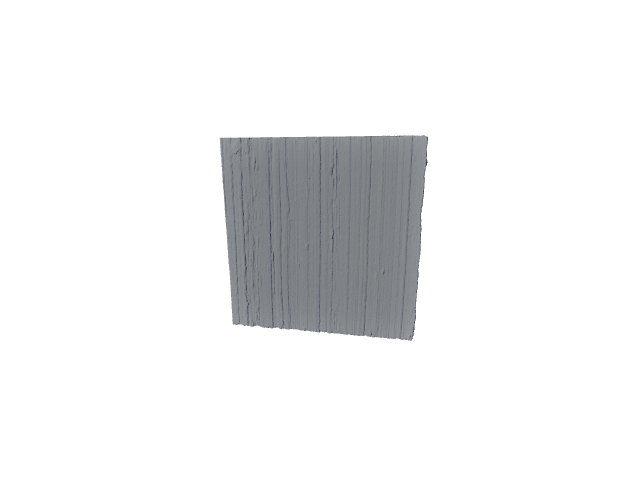} 
&
\begin{tikzpicture}
  \draw [fill=gray, fill opacity=0, dashed]
  (180:7mm) coordinate (a)
  -- ++(0,-12.5mm) coordinate (b)
  arc (180:360:7mm and 1.85mm) coordinate (d)
  -- (a -| d) coordinate (c) arc (0:180:7mm and 1.85mm);
  \draw [fill=gray, fill opacity=0]
  (0,0) coordinate (t) circle (7mm and 1.85mm);
  \draw [densely dashed] (d) arc (0:180:7mm and 1.85mm);
  \draw []
  (180:7.5mm) coordinate (A)
  -- ++(0,-12.5mm) coordinate (B) node [midway, right, inner sep=1pt] {}
  arc (180:360:7.5mm and 2.625mm) coordinate (D)
  -- (A -| D) coordinate (C) arc (0:180:7.5mm and 2.625mm);
  \draw [] (0,0) coordinate (T) circle (7.5mm and 2.625mm);
\end{tikzpicture}
&
$\xrightarrow{\hspace*{2cm}}$
&
$?$
\end{tabular}

\caption{Illustration of the problem statement.
Here the actual geometries of a small number of 3D printed steel components, each notionally a flat panel, are provided and the aim is to construct a generative statistical model for the actual geometry of a 3D-printed steel component whose notional geometry is cylindrical.
}
\label{fig: explain}
\end{figure}

\subsection{A Brief Review of Surface Metrology}

Surface metrology concerns small-scale features on surfaces and their measurement \citep{Jiang2007a,Jiang2007}. 
The principal aim is to accurately characterise the nature of the surface of a given type of material, in order that such material can be recreated in simulations or otherwise analysed.
Typical applications of surface metrology are at the nanometer-to-millimeter scale, but the principles and methods apply equally to the millimeter-to-centimeter scale variation present in 3D-printed stainless steel.

The first quantitative studies of surfaces can be traced back to the pioneering work of \cite{Abbot1933}, and concern measurement techniques to produce one-dimensional surface cross-sections (or ``profiles''), denoted $z : [0,L] \rightarrow \mathbb{R}$, and the subsequent description of profiles in terms of a small number of summary statistics, such as amplitude and texture statistics \citep{thomas1999rough}. 
Typical examples of traditional summary statistics are presented in Table \ref{tab: features from literature}. 
The use of summary statistics for surface profiles continues in modern surface metrology \citep[e.g. as codified in standard][]{ISO213}.
These include linear filters, for example based on splines and wavelets \citep{Krystek_1996}, and segmentation filters that aim to partition the profile into qualitatively distinct sections \citep{Scott_2004}.
See \cite{Jiang2007}.

\begin{table}[]
\centering
\footnotesize
	\begin{tabular}{ | l | l | p{5cm} | p{5cm} |} \hline
		\textbf{Parameter} & \textbf{Parameter} & &   \\ 
		\textbf{Name} & \textbf{Type} & \textbf{Description} & \textbf{Mathematical Definition}  \\ \hline \hline
		Rq        & Amplitude & Root mean square roughness                                                     & Rq $:= \sqrt{\frac{1}{L} \int_0^L [z(x) - \mu]^2 \mathrm{d}x}$  \\ \hline
		Ra        & Amplitude & Average roughness                                     & Ra $:= \frac{1}{L} \int_0^L |z(x) - \mu| \mathrm{d}x$          \\ \hline
		HSC       & Texture   & Number of local peaks per unit length                                   & (NA - this depends on definition of a peak)                \\ \hline
		$\text{R}_L$     & Texture   & Ratio of developed length of a profile compared to nominal length & $\text{R}_L := \frac{1}{L} \int_0^L \sqrt{1 + z'(x)^2} \mathrm{d}x$                     \\ \hline
	\end{tabular}
\caption{Summary statistics traditionally used in surface metrology; here $z : [0,L] \rightarrow \mathbb{R}$ is the profile through the surface of the material and $\mu$ is its average or intercept as per \eqref{eq: ARMA}.
Adapted from Chapters 7 and 8 of \cite{thomas1999rough}.
}
\label{tab: features from literature}
\end{table}

Although useful for the purposes of classifying materials according to their surface characteristics, these descriptive approaches do not constitute a generative model for the geometric variation in the material.
Subsequent literature in surface metrology recognised that surfaces contain inherently random features that can be characterised using a stochastic model \citep{PATIR1978263,DeVOR}. 
In particular, it has been noted that many types of rough surface are well-modelled using a non-stationary stochastic model \citep[see e.g.][]{MAJUMDAR1990313}.  
Traditional stochastic models for surface profiles fall into two main categories: The first seeks to model the surface profiles as time-series, typified by the autoregressive moving average (ARMA) model introduced in this context in \cite{DeVOR} and extended to surfaces with a non-Gaussian height distribution by \cite{WATSON1982215}. 
An ARMA$(p,q)$ model combines a $p$th order autoregressive model and a $q$th order moving average
\begin{align} \label{eq: ARMA}
z_i = \mu + \epsilon_i + \sum_{k=1}^p \phi_k z_{i-k} + \sum_{l=1}^q \theta_l \epsilon_{i-l}
\end{align}
where $z_i = z(t_i)$ represents a discretisation of the surface profile and the $\epsilon_i$ are residuals to be modelled.
These techniques have been applied to isotropic surfaces.
\cite{PATIR1978263} have noted that the theory is less well-understood regarding anisotropy and pointed out that observations of surface profiles in differing directions may be used to establish anisotropic models. 
This represents a significant weakness where simulation of rough surfaces are concerned, as it becomes unwieldly and complex to simulate materials with a layered surface such as 3D-printed steel. 
\cite{PATIR1978263} introduced a second class of methods, wherein a two-dimensional surface is represented as a $N \times M$ matrix of heights $(z_{i,j})$, assumed to have arisen as a linear transform of a $(N+n) \times (M+m)$ random matrix $(\eta_{i,j})$ according to
$$
z_{i,j} = \mu + \epsilon_i + \sum_{k=1}^{n} \sum_{l=1}^{m} \alpha_{k,l} \eta_{i+k,j+l} ,
$$
where the coefficients $(\alpha_{k,l})$ are selected in order to approximate the measured autocorrelation function for the $z_{i,j}$.
The surface heights are often assumed to be Gaussian as the theory in this setting is well-understood. 
A number of authors have developed this method:
A fractal characterisation was explored in \cite{MAJUMDAR1990313}, recognising that the randomness of some rough surfaces is multiscale and exhibits geometric self-similarity. 
\cite{HU199283} proposed the use of the fast Fourier transform on the autocorrelation function to facilitate rapid surface simulation; see also \cite{newland1980introduction} and \cite{WU200047}.
A possible drawback with this second class of approaches, compared to the former, is that it can be more difficult to account for non-stationarity compared to e.g. the ARMA model.
The use of wavelets to model non-stationary and multi-scale features has received more recent attention, see \cite{Jiang2007a}.

Recent research in surface metrology concerns the simulation of rough surfaces for applications in other fields, including tribology \citep{10.1115/1.4029644}, electronics \citep{7819933} and wave scattering \citep{CHOI201827}. 
Typically the approaches above are used directly, but alternative methods are sometimes advanced. 
For instance, \cite{10.1115/1.4029644} proposed an approach utilising random switching to generate Gaussian surfaces with prescribed statistical moments and autocorrelations. 
Clearly it is necessary to specialise any statistical model to the specific material under study.
For example, in the case of 3D-printed steel an approach ought to be taken which is able to natively account for anisotropy due to layering of the steel. 
More crucially in the present context, existing methods pre-suppose that surface profiles $(z_i)$ or height matrices $(z_{i,j})$ are pre-defined with respect to some regular Cartesian grid on a Euclidean manifold.
In applications to 3D-printed steel, where components have possibly complex macroscale geometries, the training and test data can belong to different manifolds and it is far from clear how an existing model, once fitted to training data from one manifold, could be used to simulate realistic instances of geometric variation on a different manifold.

An important application of surface metrology is \textit{surface quality monitoring}; a classification task where one seeks to determine, typically from a limited number of measurements and in an online environment, whether the quality of a manufactured component exceeds a predefined minimum level \citep{zang2018phase,zang2018phaseb,jin2019scale,zhao2020intrinsic}.
Existing research into quality monitoring has exploited Gaussian process models to reconstruct component geometries from a small number of measurements \citep{xia2008gaussian,colosimo2014profile,del2015geodesic}, possibly at different fidelities \citep{colosimo2015multisensor,ding2020multisensor}.
Going further, performance metrics arising from surface metrology  have been used as criteria against which the parameters of a manufacturing protocol can be optimised \citep{yan2019structured}.

\subsection{Outline of the Contribution}

The aim of this paper is to explore whether modern techniques from spatial statistics can be used to provide a useful characterisation of geometric variation in 3D-printed stainless steel.

\begin{figure}[t!]
\centering
\begin{subfigure}[b]{0.29\textwidth}
\vspace{-20pt}
\includegraphics[width = \textwidth]{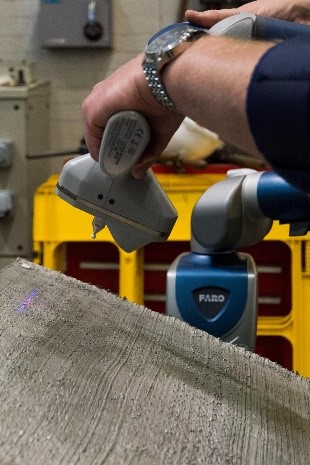}

\vspace{15pt}
\caption{}
\end{subfigure}
\begin{subfigure}[b]{0.6\textwidth}
\scalebox{.5}{
\includegraphics[clip,trim = 8cm 13cm 9.3cm 0cm]{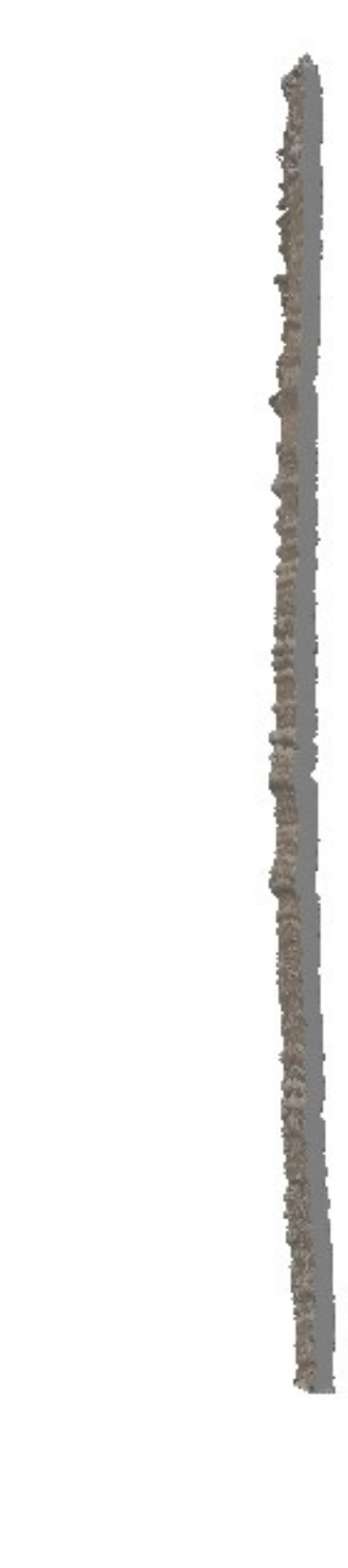}
\includegraphics[clip,trim = 4cm 13cm 3cm 0cm]{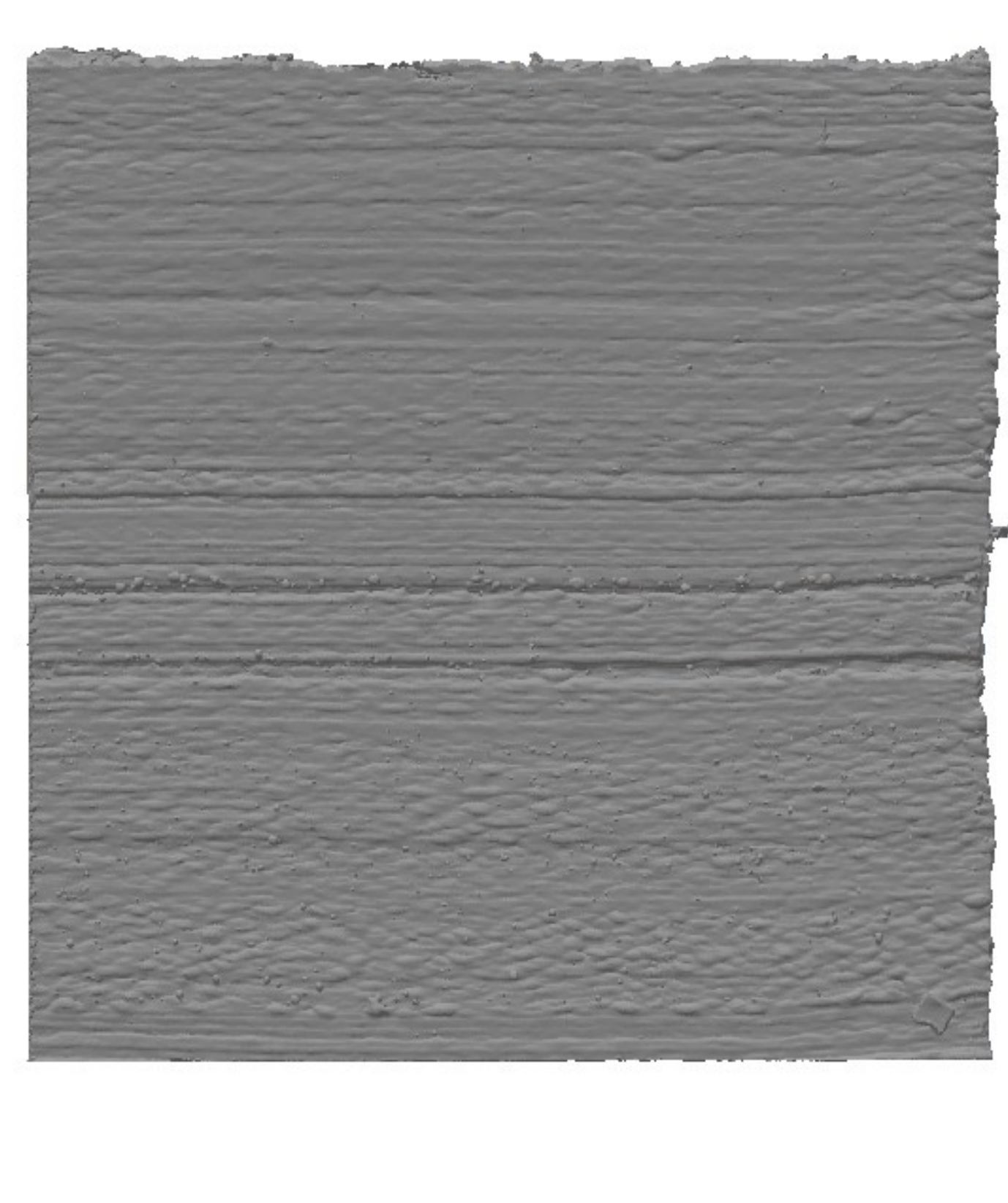}
}

\vspace{-30pt}
\scalebox{.5}{
\phantom{\includegraphics[clip,trim = 8cm 30cm 9.3cm 0cm]{figures/35_panel_side.pdf}}
\includegraphics[clip,trim = 4cm 21cm 3cm 7cm]{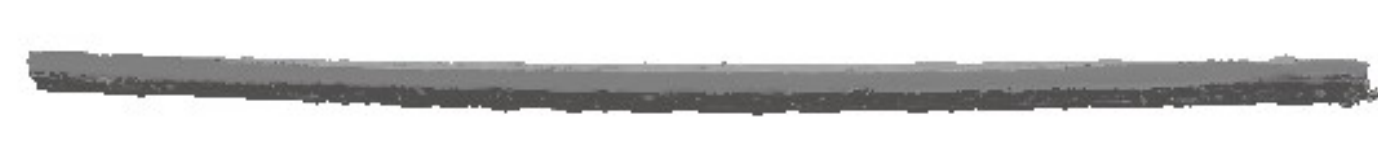}
}
\caption{}
\label{fig: example panel}
\end{subfigure}
\caption{Laser scan of 3D-printed steel sheet.
(a) Photograph of the handheld scanning equipment.
(b) Orthographic projection of (a portion of) the scanned sheet, which will be called a \emph{panel}. 
The notional thickness of the panel is 3.5mm.
[Observe the residual stress induces a slight curvature in the notionally flat panel.] }
\label{fig: protocol}
\end{figure}

To provide a training dataset, a high-resolution laser was used to scan 3D-printed stainless steel panels, on whose surfaces the geometric variation of interest is exhibited.
A typical panel is displayed in Figure \ref{fig: protocol}.
To characterise geometric variation in the panel dataset, in Section \ref{sec: methods} a library of candidate models is constructed, rooted in Gaussian stochastic processes, and in Section \ref{sec: results} formal parameter inference and model selection are performed to identify a most suitable candidate model.
Although quite natural from a statistical perspective, this systematic approach to parameter inference and model selection appears to be novel in the surface metrology context.
The high-dimensional nature of the dataset introduces computational challenges that are familiar in spatial statistics and the stochastic partial differential equation (SPDE) techniques described in \cite{Lindgren2011} are leveraged to make the computation practical.

The aim is to characterise geometric variation not just on panels but on components whose notional geometry, represented abstractly by a two-dimensional manifold $\mathcal{M}$ embedded in $\mathbb{R}^3$, may be complicated. 
(The case study in \Cref{subsec: predict cylinder 1} considers the case where $\mathcal{M}$ is cylindrical which, whilst not complicated, differs from a flat panel in a fundamental topological way.)
This variation, present in the actual geometry of a 3D-printed component, is described by a two-dimensional vector field $u : \mathcal{M} \rightarrow \mathbb{R}^2$.
As illustrated in Figure \ref{fig: notation illustration}, the value of the first coordinate $u^{(1)}(x)$ of the vector field represents the Euclidean distance of the upper surface of the component from the manifold and similarly the value of $u^{(2)}(x)$ represents the distance of the lower surface of the component from the manifold. 
The absence of geometric variation corresponds to the functions $u^{(1)}(x)$ and $u^{(2)}(x)$ being constant, but actual 3D-printed steel components exhibit geometric variation that can be statistically modelled.
The need to consider diverse notional geometries, represented by different manifolds, precludes conventional parametric \citep{Rasmussen2006}, compositional \citep{Duvenaud2014} and nonparametric \citep{Bazavan2012,Oliva2016,Moeller2016} approaches that aim to infer an appropriate covariance function defined only on the manifold of the training dataset.
Indeed, it is well-known that a radial covariance function $\varphi(d_{\mathcal{M}_1}(x,y))$ defined on one manifold $x,y \in \mathcal{M}_1$ with geodesic distance $d_{\mathcal{M}_1}$ can fail to be a valid covariance function when applied as $\varphi(d_{\mathcal{M}_2}(x,y))$ in the context of a different manifold $x,y \in \mathcal{M}_2$ with geodesic distance $d_{\mathcal{M}_2}$.
This occurs because positive-definiteness can fail; see Section 4.5 of \cite{Gneiting_2013}.
The key insight used in this paper to generalise across manifolds is that the Laplacian $\Delta_{\mathcal{M}_1}$ on $\mathcal{M}_1$ can be associated with the Laplacian $\Delta_{\mathcal{M}_2}$ on $\mathcal{M}_2$.
In doing so a suitable differential operator on $\mathcal{M}_1$, describing the statistical character of the vector field $u$ via an SPDE, can be inferred and then used to instantiate an equivalent differential operator on $\mathcal{M}_2$, defining a Gaussian random field on $\mathcal{M}_2$ via an SPDE. 
This allows, for instance, the simulation of high-resolution random surfaces on a cylinder whose local characteristics are related to the panel training dataset.
(Note that this paper does not consider the question of when the covariance differential operator / SPDE approach defines a true random field rather than a generalised random field.
The practical consequences of this are minimal, however, since the random field will be discretised by projection onto a finite element basis whose coefficients are well-defined linear functionals of the random field.)

In Sections \ref{subsec: predict cylinder 1} and \ref{subsec: predict cylinder 2} it is demonstrated, using a held out test dataset consisting of a 3D-printed steel cylinder, that random surfaces generated in this way can capture some salient aspects of geometric variation in the material.
However, the conclusions of this paper are tempered by the substantial computational challenges associated with the fitting of model parameters in this context.
The paper concludes with a brief discussion in Section \ref{sec: conclusion}.

\begin{figure}[t!]
\centering
\includegraphics[width = 0.9\textwidth,clip,trim = 2cm 3cm 0cm 1cm]{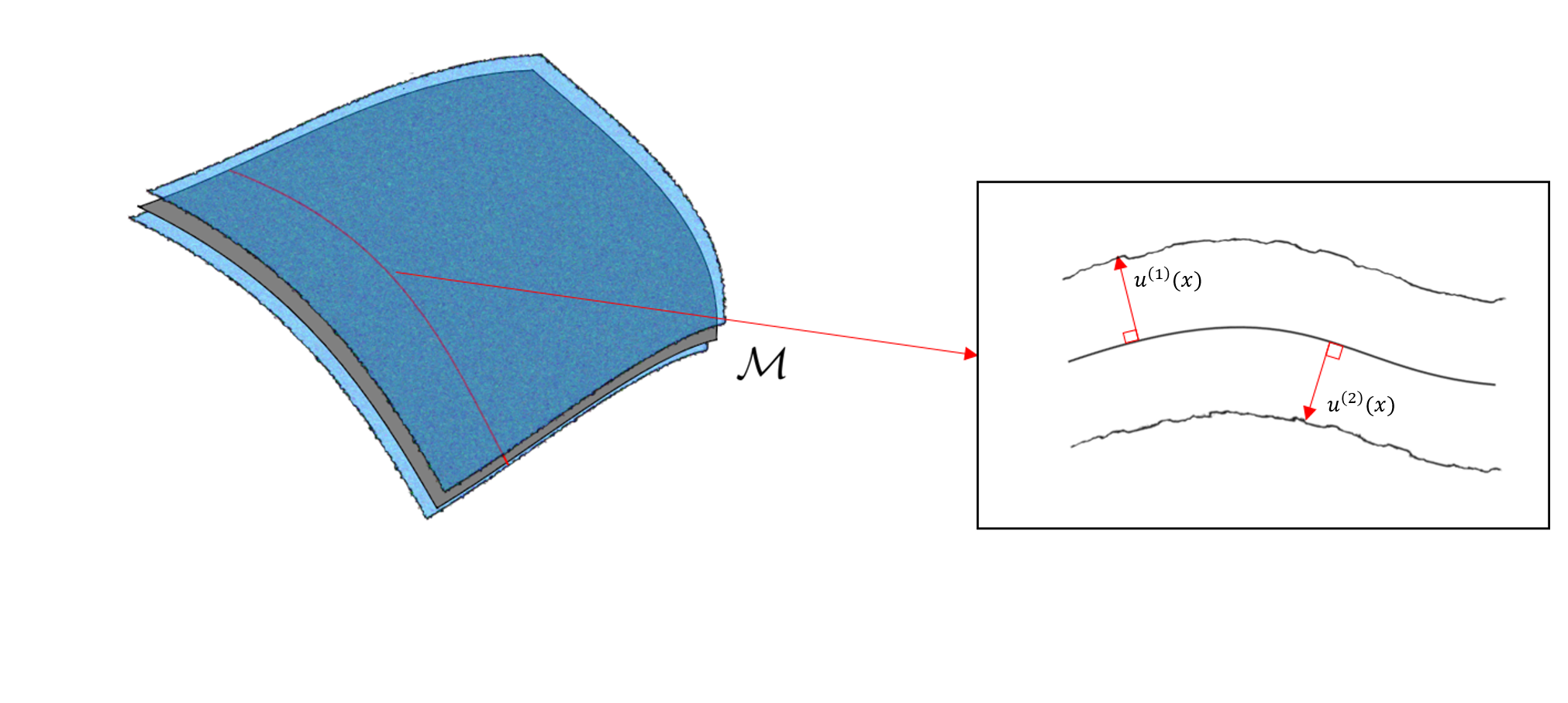}
\caption{Data are conceptualised as a two-dimensional vector field $u$ on an oriented manifold $\mathcal{M}$.
The value $u^{(1)}(x)$ represents the height of the upper surface of the material above $x \in \mathcal{M}$, while the value $u^{(2)}(x)$ represents the height of the lower surface below $x \in \mathcal{M}$.
}
\label{fig: notation illustration}
\end{figure}

This research forms part of a wider effort to develop realistic computer modelling tools for applications involving 3D-printed steel; these must also account also for other factors such as mechanical variation and aspects of the printing protocol \citep{Dodwell2021}.
Specifically, other controllable factors that affect the quality of the manufactured component include the angle of the weld head, the speed of printing and the order in which the steel layers are printed.
It is possible that an improved understanding of the nature of geometric variation can inform the improvement of manufacturing techniques and their protocols \citep{Barclift2012,Franco2017}, but these directions are beyond the scope of the present paper, which aims only to characterise the geometric variation in the material.

\section{Statistical Models for Geometric Variation in 3D-Printed Steel} \label{sec: methods}

This section proceeds as follows:
Section \ref{subsec: insight} contains the main conceptual contribution, that identification of Laplacians can be used to generalise from the training manifold to a test manifold.
Section \ref{subsec: maths} presents the mathematical setting of a two-dimensional Gaussian random field characterised by a differential operator and Section \ref{subsec: candidates} presents the resulting collection of candidate models.

\subsection{Gaussian Models and Generalisation via the Laplacian} \label{subsec: insight}

In this paper, geometric variation for a 3D-printed component with notional geometry $\mathcal{M}$ is modelled using a Gaussian stochastic process.
Letting $\mathbb{E}$ and $\mathbb{C}$ denote expectation and covariance operators, set
\begin{align*}
\mathbb{E}[u^{(r)}(x)] & = 0 \\
\mathbb{C}( u^{(r)}(x) , u^{(s)}(y) ) & = k_{\mathcal{M}}((x,r),(y,s)) ,
\end{align*}
where the covariance function $k_{\mathcal{M}}((x,r),(y,s))$ describes the statistical relationship between the upper and lower surfaces ($r,s \in \{1,2\}$) at different spatial locations ($x,y \in \mathcal{M}$).
To describe geometric variation in the panel datasets, whose notional flat geometry is denoted $\mathcal{M}_1$, any existing method for estimating a suitable covariance function $k_{\mathcal{M}_1}$ from the dataset could be employed.
Such methods are well-developed and examples include maximum likelihood estimation over a parametrised class of covariance functions \citep{Rasmussen2006}, inference based on cross-validation with arbitrarily complicated compositions of simple kernels \citep{Duvenaud2014} and even fully nonparametric inference of a covariance function \citep{Bazavan2012,Oliva2016,Moeller2016}.
However, in doing so one would be learning a covariance function that is defined on the manifold $\mathcal{M}_1$ of the training dataset, being the Green's function of a stochastic differential equation defined on $\mathcal{M}_1$ \citep[see e.g.][]{Fasshauer2013}. 
As a consequence, one cannot expect a covariance function learned from the training dataset to carry the same meaning when applied on the manifold $\mathcal{M}_2$ of the test dataset. 
Indeed, it is a classic result that the Mat\'{e}rn kernel $\varphi$ on $\mathbb{R}^3$ with $d_{\mathbb{R}^3}(x,y) = \|x-y\|$ can fail to be positive definite when applied as $\varphi(d_{\mathbb{S}^2}(x,y))$ on the sphere $x,y \in \mathbb{S}^2 \subset \mathbb{R}^3$ with $d_{\mathbb{S}^2}$ the geodesic distance on $\mathbb{S}^2$ \citep[see e.g. Section 4.5 of][]{Gneiting_2013}.
Thus inferential approaches based on the covariance function are unsuitable in this context, since a description is required of geometric variation on a range of manifolds that includes potentially complicated manufacturing geometries.

The key insight in this paper is that, whilst covariance functions do not generalise across manifolds, their associated differential operators are only locally defined. 
In particular, a wide range of models for stochastic vector fields $u$ can be cast as the solutions of a coupled system of SPDEs based on a Laplacian-type differential operator.
The specific form of the differential operator, together with suitable boundary conditions, determine the statistical properties of the associated random field.
The Laplacian $\Delta_{\mathcal{M}_1}$ on the manifold $\mathcal{M}_1$ of the training dataset can be associated with the Laplacian $\Delta_{\mathcal{M}_2}$ on the manifold $\mathcal{M}_2$ where generative models are to be tested. 
This provides a natural way in which to decouple the statistical properties of the random vector field $u$ from the manifold on which it is defined.
To sound a cautionary note, this assumption is likely to break down in situations where one of the manifolds is highly curved.
For example, it is conceivable that the protocols used to control weld head movement may be less effective when the required trajectories are highly curved, leading to higher levels of geometric variation in corresponding regions of the component.
As a more extreme example, if two distinct regions of the manifold come within millimetres due to the manifold curving back on itself when embedded in 3D, then through stochastic fluctuation there is a potential for the two corresponding regions of the printed component to become physically fused.
The first of these phenomena could be handled by considering Gaussian models that allow covariances to be both position- and curvature-dependent, while an entirely different approach would be needed to model accidental fusing of distinct regions of a component.
Neither of these extensions are pursued in the present work.

The SPDE formulation of Gaussian random fields has received considerable attention in the spatial statistics literature following the seminal work of \cite{Lindgren2011}, generating an efficient computational toolkit which has been widely used \citep[e.g.][]{coveney2019probabilistic}.
Interestingly, the present account appears to describe the first instance of \emph{transfer learning} using the SPDE framework, where observations made on one manifold are used to provide information on the equivalent stochastic process defined on a different manifold.
As noted earlier, no attempt is made to address the technical issue of whether the covariance operator corresponds to a true random field rather than a generalised random field.

\subsection{Mathematical Set-Up} \label{subsec: maths}

The purpose of this section is to rigorously set up a coupled SPDE model for a random vector field $u$ on a manifold $\mathcal{M}$.
For completeness, the required regularity assumptions are stated in Section \ref{subsubsec: reg}; these are essentially identical to the presentation in \cite{Lindgren2011} and may be safely skipped if desired.
The coupled SPDE model is introduced in Section \ref{subsubsec: TDGRF}, following a similar presentation to that of \cite{Hu2013a,Hu2013}.

\subsubsection{Regularity Assumptions} \label{subsubsec: reg}

The domain of the printing process is $\mathbb{R}^3$ and throughout the paper the convention is used that the printing occurs in the direction of the $z$-axis of $\mathbb{R}^3$. 
That is, each layer of steel is notionally perpendicular to the $z$-axis according to the printing protocol.
The notional geometry of a component is described by a manifold $\mathcal{M}$ embedded in $\mathbb{R}^3$.
Below are listed some strong regularity conditions on $\mathcal{M}$ that ensure the coupled SPDE model in Section \ref{subsubsec: TDGRF} is well-defined. 

Let $\mathcal{M}$ be a 2-dimensional manifold embedded in $\mathbb{R}^3$, so that the Hausdorff measure on $\mathcal{M}$, denoted $\mu_\mathcal{M}$, can be defined, and thus an inner product
\begin{align*}
\langle \phi , \psi \rangle_\mathcal{M} & := \int_\mathcal{M} \phi(x) \psi(x) \mathrm{d} \mu_\mathcal{M}(x)
\end{align*}
and an associated space $L^2(\mathcal{M})$ of functions $\phi : \mathcal{M} \rightarrow \mathbb{R}$ for which $\langle \phi , \phi \rangle_\mathcal{M} < \infty$. 
It is assumed that $\mathcal{M}$ is endowed with a Riemannian structure, induced by the embedding in $\mathbb{R}^3$, which implies that one can define differential operators including the gradient $\nabla_{\mathcal{M}}$, the Laplacian $\Delta_{\mathcal{M}}$ and the normal derivative $\partial_{\mathcal{M},\mathrm{n}}$ on $\mathcal{M}$. 
It is assumed that $\mathcal{M}$ is compact (with respect to the topology induced by $\mathbb{R}^3$), which implies the existence of a countable subset $\{E_k : k = 0,1,\dots\}$ of eigenfunctions of the negative Laplacian, $- \Delta_{\mathcal{M}} E_k = \lambda_k E_k$, which can be taken as a basis for $L^2(\mathcal{M})$.
From compactness it follows that $\mathcal{M}$ is bounded and it is further assumed that $\mathcal{M}$ has a boundary $\partial \mathcal{M}$ that is a piecewise smooth 1-dimensional manifold. 
(From a practical perspective, the assumption that the manifold has a boundary is not restrictive, since manifolds without boundary, such as the sphere, are likely to be challenging to realise as 3D-printed components.)
The induced measure on $\partial \mathcal{M}$ is denoted $\mu_{\partial \mathcal{M}}$.
In the sequel the manifold $\mathcal{M}$ is fixed and the shorthand $\Delta = \Delta_{\mathcal{M}}$ is used, leaving the dependence of the Laplacian on $\mathcal{M}$ implicit.

\subsubsection{Two-Dimensional Gaussian Random Fields} \label{subsubsec: TDGRF}

\cite{Hu2013a,Hu2013} considered using coupled systems of SPDEs to model random vector fields, building on the constructions in \cite{Lindgren2011}.
In this work a similar coupled SPDE construction is used, defined next.

Let $\Omega$ be an underlying probability space, on which all random variables in this paper are defined, and let $\omega \mapsto Z^{(r)}(x; \omega)$, $r \in \{1,2\}$ be independent, centred, real-valued (true or generalised) random fields, whose distribution is to be specified.
Following standard convention, the argument $\omega \in \Omega$ is left implicit throughout. 
Let $\mathcal{L}^{(rs)}$, $r,s \in \{1,2\}$, be differential operators on $\mathcal{M}$, to be specified, and consider then the coupled system of SPDEs
\begin{align}
\left[ \begin{array}{cc} \mathcal{L}^{(11)} & \mathcal{L}^{(12)} \\ \mathcal{L}^{(21)} & \mathcal{L}^{(22)} \end{array} \right] \left[ \begin{array}{c} u^{(1)}(x) \\ u^{(2)}(x) \end{array} \right] = \left[ \begin{array}{c} Z^{(1)}(x) \\ Z^{(2)}(x) \end{array} \right] , \qquad x \in \mathcal{M} , \label{eq: SPDE1} 
\end{align}
whose specification is completed with Neumann boundary conditions on $\partial \mathcal{M}$.
The system in \eqref{eq: SPDE1} is compactly represented in matrix form as $\mathcal{L} \mathbf{u} = \mathbf{Z}$.
The distribution of $u$, the vector field solution of \eqref{eq: SPDE1}, is therefore fully determined by the specification of the matrix differential operator $\mathcal{L}$ and the distribution of the random field $\mathbf{Z}$.
In particular, the components $u^{(r)}$, $r \in \{1,2\}$ are independent if $\mathcal{L}^{(12)} = \mathcal{L}^{(21)} = 0$ and if $Z^{(1)}$ and $Z^{(2)}$ are independent.
Suitable choices for $\mathcal{L}$ and $\mathbf{Z}$ are discussed next.

\subsection{Candidate Models}  \label{subsec: candidates}

A complete statistical description for the random vector field $u$ follows from specification of both the differential operator $\mathcal{L}$ and the noise process $\mathbf{Z}$.
Sections \ref{subsubsec: iso model}-\ref{subsubsec: ani nonstat} present four candidate parametric models for $\mathcal{L}$, then Sections \ref{subsubsec: white}-\ref{subsubsec: smo osc noise} present three candidate parametric models for $\mathbf{Z}$.
Thus a total of 12 candidate parametric models are considered for the random vector field $u$ of interest. 

It is important to acknowledge that the SPDE framework does not currently enjoy the same level of flexibility for modelling of complex structural features as the covariance function framework.
This is for two reasons; first, the indirect relationship between an SPDE and its associated random field presents a barrier to the identification of a suitable differential operator to represent a particular feature of interest and, second, each SPDE model requires the identification or development of a suitable numerical (e.g. finite element) method in order to facilitate computations based on the model.
A discussion of the numerical methods used to discretise the SPDEs in this work is reserved for Appendix A.1, with model-specific details contained in Appendix A.2.
For these two reasons, the aim in the sequel is not to arrive at a generative model whose realisations are indistinguishable from the real samples of the material. Rather, the aim is to arrive at a generative model whose realisations are somewhat realistic in terms of the salient statistical properties that determine performance in the engineering context. 
The models were therefore selected to exhibit a selection of physically relevant features, including anisotropy, oscillatory behaviour, degrees of smoothness and non-stationarity of the random field.
It is hoped that this paper will help to catalyse collaborations with the surface metrology community that will lead in the future to richer and more flexible SPDE models for geometric variation in material.

The candidate models presented next are similar to those considered in \cite{Lindgren2011}, but generalised to coupled systems of SPDEs in a similar manner to \cite{Hu2013,Hu2013a}. 
Note that formal model selection was not considered in \cite{Lindgren2011} or \cite{Hu2013a,Hu2013}.

\subsubsection{Isotropic Stationary} \label{subsubsec: iso model}

For the differential operators $\mathcal{L}^{(rs)}$, $r,s \in \{1,2\}$, the simplest candidate model that considered is 
\begin{align}
\mathcal{L}^{(rs)} u(x) := (\eta^{(rs)} - \Delta) ( \tau^{(rs)} u(x) ) , \qquad x \in \mathcal{M} , \label{eq: iso stat L}
\end{align}
where $\eta^{(rs)}, \tau^{(rs)} > 0$ are constants to be specified. 
This model is stationary, meaning that the differential operator $\mathcal{L}^{(rs)}$ does not depend on $x$, and isotropic, meaning that it is invariant to rotation of the coordinate system on $\mathbb{R}^3$.
The differential order of the differential operator in \eqref{eq: iso stat L} is fixed, which can be contrasted with \cite{Lindgren2011} where powers $(\eta - \Delta)^{\alpha/2}$, $\alpha \in \mathbb{N}$, of the pseudo-differential operator were considered.
The decision to fix $\alpha = 2$ was taken because (a) the roughest setting of $\alpha = 1$ was deemed not to be realistic for 3D-printed steel, since the solutions $u$ to the SPDE are then not well-defined, and (b) over-parametrisation will be avoided by controlling smoothness of the vector field $u$ through the smoothness of the driving noise process $\mathbf{Z}$, rather than through the differential order of $\mathcal{L}$. 
The model parameters in this case are therefore $\theta = \{\tau^{(rs)},\eta^{(rs)}\}_{r,s \in \{1,2\}}$.
To reduce the degrees of freedom, it was assumed that $\eta^{(12)} = \eta^{(21)}$, $\eta^{(11)} = \eta^{(22)}$, $\tau^{(12)} = \tau^{(21)}$ and $\tau^{(11)} = \tau^{(22)}$. 
This encodes an exchangeability assumption on the statistical properties of the upper and lower surfaces of the material.
Such an assumption seems reasonable since the distinction between upper and lower surface was introduced only to improve the pedagogy; in reality such a distinction is arbitrary.

\subsubsection{Anisotropic Stationary} \label{subsubsec: aniso model}

The visible banding structure in Figure \ref{fig: example panel}, formed by the sequential deposition of steel as the material is printed, suggests that an isotropic model is not well-suited to describe the dataset.
To allow for the possibility that profiles parallel and perpendicular to the weld bands admit different statistical descriptions, the following anisotropic stationary model was considered: 
\begin{align}
\mathcal{L}^{(rs)}u(x) := (\eta^{(rs)} - \nabla \cdot \mathbf{H}^{(rs)} \nabla)(\tau^{(rs)} u(x)), \qquad x \in \mathcal{M} .  \label{eq: aniso stat pde}
\end{align}
Here \eqref{eq: aniso stat pde} differs from \eqref{eq: iso stat L} through the inclusion of a diffusion tensor $\mathbf{H}^{(rs)}$, $r,s \in \{1,2\}$, a positive definite $3 \times 3$ matrix whose elements determine the nature and extent of the anisotropy being modelled.
The model parameters in this case are $\theta = \{\eta^{(rs)},\tau^{(rs)}, \mathbf{H}^{(rs)} \}_{r,s \in \{1,2\}}$.
To reduce the number of parameters, it was assumed that $\mathbf{H}^{(rs)} \equiv \mathbf{H}$ where $\mathbf{H}$ is a $3 \times 3$ diagonal matrix, so that the nature of the anisotropy is the same for each of the four differential operators being modelled.
Moreover, anisotropy was entertained only in the direction of printing (the $z$-axis of $\mathbb{R}^3$) relative to the non-printing directions, meaning that the model is in fact \emph{orthotropic} with $\mathbf{H} = \text{diag}(h_1,h_1,h_2)$ for as yet unspecified $h_1,h_2 > 0$.
As before, an exchangeability assumption on the upper and lower surfaces of the material was enforced, so that $\eta^{(12)} = \eta^{(21)}$, $\eta^{(11)} = \eta^{(22)}$, $\tau^{(12)} = \tau^{(21)}$ and $\tau^{(11)} = \tau^{(22)}$.

\subsubsection{Isotropic Non-Stationary} \label{subsubsec: nonstat model}

A more detailed inspection of the material in Figure \ref{fig: example panel} reveals that (at least on the visible surface) the vertical centre of the panel exhibits greater variability in terms of surface height compared to the other portions of the panel.
It can therefore be conjectured that the printing process is non-stationary with respect to the direction of the $z$-axis, in which case this non-stationarity should be encoded into the stochastic model.
To entertain this possibility, a differential operator of the form
\begin{align*}
\mathcal{L}^{(rs)} u(x) & := (\eta^{(rs)}(x_3) - \Delta) (\tau^{(rs)}(x_3) u(x) ), \qquad x \in \mathcal{M} ,
\end{align*}
was considered, which generalises \eqref{eq: iso stat L} by allowing the coefficients $\eta^{(rs)}$ and $\tau^{(rs)}$ to depend on the $z$-coordinate $x_3$ of the current location $x$ on the manifold. 
Here it was assumed that $\tau^{(rs)}(x_3) = c_\tau^{(rs)} \tau(x_3)$ and $\eta^{(rs)}(x_3) = c_\eta^{(rs)} \eta(x_3)$ for some unspecified $c_\tau^{(rs)}$, $c_\eta^{(rs)}$ and some functions $\tau$, $\eta$, to be specified, whose scale is fixed. 
The nature of the non-stationarity, as characterised by $\tau$ and $\eta$, will differ from printed component to printed component and this higher order variation could be captured by a hierarchical model that can be considered to have given rise to each specific instance of the $\tau,\eta$.

Recent work due to \cite{Roininen2016} considered the construction of non-stationary random fields that amounted to using a nonparametric Gaussian random field for both $\tau$ and $\eta$.
However, this is associated with a high computational overhead that prevents one from deploying such an approach at scale in the model selection context. 
Therefore the functions $\tau$ and $\eta$ were instead parametrised using a low order Fourier basis, for $\tau$ taking
\begin{equation}
\tau(x) = 1 + \sum_{i=1}^4 \sum_{j=0}^1 \gamma_\tau^{(ij)} \sin \left( \frac{ 2 \pi i } {100} x_3 + \frac{j \pi}{4} \right) \label{eq: nonatationary param}
\end{equation}
where the $\gamma_\tau^{(\cdot)}$ are parameters to be specified. 
The units of $x_3$ are millimeters, so \eqref{eq: nonatationary param} represents low frequency non-stationarity in the differential operator.
For $\eta$ an identical construction was used, with coefficients denoted $\gamma_\eta^{(\cdot)}$.
The model parameters in this case are $\theta = \{c_\tau^{(rs)}, c_\eta^{(rs)},  \gamma_\tau^{(ij)}, \gamma_\eta^{(ij)} \}$.
To reduce the number of parameters an exchangeability assumption was again used, setting $c_\tau^{(11)} = c_\tau^{(22)}$, $c_\tau^{(12)} = c_\tau^{(21)}$, $c_\eta^{(11)} = c_\eta^{(22)}$ and $c_\eta^{(12)} = c_\eta^{(21)}$. 

To circumvent the prohibitive computational cost of fitting a hierarchical model for the Fourier coefficients $\gamma_\tau^{(\cdot)}$ and $\gamma_\eta^{(\cdot)}$, joint learning of these parameters across the panel datasets was not attempted.
This can be justified by the fact that we have only 6 independent samples, drastically limiting the information available for any attempt to learn a hierarchical model from the dataset.

\subsubsection{Anisotropic Non-stationary} \label{subsubsec: ani nonstat}

The final model considered for the differential operator is the natural combination of the anisotropic model of Section \ref{subsubsec: aniso model} and the non-stationary model of Section \ref{subsubsec: nonstat model}.
Specifically, the differential operator
\begin{align*}
\mathcal{L}^{(rs)} u(x) := (\eta^{(rs)}(x_3) - \nabla \cdot \mathbf{H}^{(rs)} \nabla)(\tau^{(rs)}(x_3) u(x) )  , \qquad x \in \mathcal{M}  
\end{align*}
was considered, where the $3 \times 3$ matrices $\mathbf{H}^{(rs)}$ and the functions $\tau^{(rs)}$, $\eta^{(rs)}$ are parametrised in the same manner, with the same exchangeability assumptions, as earlier described. 
Note that the matrices $\mathbf{H}^{(rs)}$ are not considered to be spatially dependent in this model.

\subsubsection{White Noise} \label{subsubsec: white}

Next, attention turns to the specification of a generative model for the noise process $\mathbf{Z}$. 
In what follows three candidate models are considered, each arising in some way from the standard white noise model.
The simplest of these models is just the white noise model itself, described first.

Let $Z^{(j)} := W^{(r)}$ be a spatial random white noise process on $\mathcal{M}$, defined as an $L^2(\mathcal{M})$-bounded generalised Gaussian field such that, for any test functions $\{\varphi_i \in L^2(\mathcal{M}) : i = 1,\dots,n\}$, the integrals $\langle \varphi_i , W^{(j)} \rangle_\mathcal{M}$, $i = 1, \dots, n$, are jointly Gaussian with 
\begin{align}
\mathbb{E} \langle \varphi_i , W^{(j)} \rangle_\mathcal{M} & = 0 \nonumber \\
\mathbb{C}( \langle \varphi_i , W^{(j)} \rangle_\mathcal{M} , \langle \varphi_j , W^{(j)} \rangle_\mathcal{M} ) & = \langle \varphi_i , \varphi_j \rangle_\mathcal{M} . \label{eq: White def 2}
\end{align}
It is not necessary to introduce a scale parameter for the noise process since the $\tau^{(rs)}$ parameters perform this role in the SPDE.

For this choice of noise model the random field $u$ of interest is typically continuous but not mean-square differentiable, which may or (more likely) may not be an appropriate mathematical description of the material.

\subsubsection{Smoother Noise} \label{subsubsec: smoother noise}

It is difficult to determine an appropriate level of smoothness for the random field $u$ based on visual inspection, except to rule out models that cannot be physically realised (c.f. Section \ref{subsubsec: iso model}).
For this reason, a smoother model for the noise process $\mathbf{Z}$ was also considered. 
In this case the noise process is itself the solution of an SPDE
\begin{align}
(\eta^{(r)} - \nabla \cdot \mathbf{H} \nabla) Z^{(r)}(x) & = W^{(r)}(x) & x \in \mathcal{M} \label{eq: smooth noise pde} \\
\partial_{\mathrm{n}} Z^{(r)}(x) & = 0 & x \in \partial\mathcal{M} \nonumber
\end{align}
where $W^{(r)}$ is the white noise process defined in Section \ref{subsubsec: white}.
To limit scope, it was assumed that the matrix $\mathbf{H}$ is the same as the matrix used (if indeed one is used) in the construction of the differential operator in Sections \ref{subsubsec: aniso model} and \ref{subsubsec: ani nonstat}.
The only free parameters are therefore $\eta^{(1)}, \eta^{(2)} > 0$ and the further exchangeability assumption was made that these two parameters are equal. 

For this choice of noise model the random field $u$ of interest is typically twice differentable in the mean square sense, which may or may not be more realistic than the white noise model.

\subsubsection{Smooth Oscillatory Noise} \label{subsubsec: smo osc noise}

The final noise model allows for the possibility of oscillations in the realisations of $\mathbf{Z}$. 
As explained in Appendix C4 of \cite{Lindgren2011}, such noise can be formulated as arising from the complex SPDE
\begin{align*}
(\eta e^{i \pi \theta} - \nabla \cdot \mathbf{H} \nabla ) [ Z^{(1)}(x) + i Z^{(2)}(x) ] & = W^{(1)}(x) + i W^{(2)}(x) & x \in \mathcal{M} \\
\partial_n [Z ^{(1)}(x) + i Z^{(2)}(x) ] & = 0 & x \in \partial\mathcal{M} . 
\end{align*}
Here $\eta > 0$ and $0 \leq \theta < 1$ are to be specified, with $\theta = 0$ corresponding to the smoother noise model of \Cref{subsubsec: smoother noise}.
The matrix $\mathbf{H}$ is again assumed to be equal to the corresponding matrix used (if indeed one is used) in the construction of the differential operator in Sections \ref{subsubsec: aniso model} and \ref{subsubsec: ani nonstat}, and $W^{(1)}$ and $W^{(2)}$ are independent white noise processes as defined in Section \ref{subsubsec: white}.

This completes the specification of candidate models for $\mathcal{L}$ and $\mathbf{Z}$, constituting 12 different models for the random vector field $u$ in total. 
These will be denoted $M_i$, $i = 1,\dots,12$ in the sequel.
Section \ref{sec: results} seeks to identify which of these models represents the best description of the panel training dataset.

\section{Data Pre-Processing and Model Selection} \label{sec: results}

This section proceeds as follows:
In Section \ref{subsec: set up} the training dataset is described and Section \ref{subsec: preprocess} explains how it was pre-processed.
Section \ref{subsec: model select} describes how formal statistical model selection was performed to identify a ``best'' model from the collection of 12 models just described.
The fitting of model parameters raises substantial computational challenges, which this paper does not fully solve but discusses in detail.

\subsection{Experimental Set-Up} \label{subsec: set up}

The data were obtained from large sheets of notional 3.5mm thickness manufactured by the Dutch company MX3D (\url{https://mx3d.com}) using their proprietary printing protocol. 
The dataset consisted of 6 panels of approximate dimensions 300mm $\times$ 300mm, cut from a large sheet.
These panels were digitised using a laser scanner to form a data point cloud; the experimental set-up and a typical dataset are displayed in Figure \ref{fig: protocol}.
The measuring process itself can be considered to be noiseless, due to the relatively high (0.1mm) resolution of the scanning equipment.
For each panel a dense data point cloud was obtained, capturing the geometry of both sides of the panel.

\subsection{Data Pre-Processing} \label{subsec: preprocess}

Before data were analysed they were pre-processed and the steps taken are now briefly described.

\subsubsection{Cropping of the Boundary}

First, as is clear from Figure \ref{fig: example panel}, part of the boundary of the panel demonstrates a high level of variation. 
This variation is not of particular interest, since in applications the edges of a component can easily be smoothed.
Each panel in the dataset was therefore digitally cropped using a regular square window to ensure that boundary variation was removed.
Note that no reference coordinate system was available and therefore the orientation of the square window needed to be determined (i.e. a \textit{registration} task needed to be solved).
To identify the direction perpendicular to the plane of the panel a principal component analysis of the data point cloud was used to determine the direction of least variation; this was taken to be the direction perpendicular to the plane of the panel.
To orient the square window in the plane a coordinate system was chosen to maximise alignment with the direction of the weld bands.
The weld bands are clearly visible (see Figure \ref{fig: example panel}) and the average intensity of a Gabor filter was used to determine when maximal agreement had been reached \citep[see e.g.][]{Weldon1996}.

\subsubsection{Global De-Trending}

As a weld cools, a contraction in volume of the material introduces \emph{residual stress} in the large sheet \citep{Radaj2012}.
As a consequence, when panels are cut from the large sheet these stresses are no longer contained, resulting in a global curvature in each panel as internal forces are re-equilibrated.
This phenomenon manifests as a slight curvature in the orthographic projections of the panel shown in Figure \ref{fig: example panel}.
The limited size of the dataset, in terms of the number of panels cut from the large sheet, precludes a detailed study of geometric variation due to residual stress in this work.
Therefore residual stresses were removed from each panel dataset by subtracting a quadratic trend, fitted using a least-squares objective in the direction perpendicular to the plane of the panel.
This narrows focus to the variation that is associated with imprecision in the motion of the print head as steel is deposited.
The data at this stage constitute a point cloud in $\mathbb{R}^3$ that is centred around a two-dimensional, approximately square manifold $\mathcal{M}$, aligned to the $x$ and $y$ axes in $\mathbb{R}^3$.

\subsubsection{High-Frequency Filter}

Closer inspection of Figure \ref{fig: example panel} reveals small bumps on the surface of the panel.
These are referred to as ``splatter'' and result from small amounts of molten steel being sprayed from the weld head as subsequent layers of the sheet are printed.
Splatter is not expected to contribute much to the performance of the material and therefore is of little interest.
To circumvent the need to build a statistical model for splatter, a pre-processing step was used in order to remove such instances from the dataset.
For this purpose a two-dimensional Fourier transform was applied independently to each surface of each panel and the high frequencies corresponding to the splatter were removed.
Since data take the form of a point cloud, the data were first projected onto a regular Cartesian grid, with grid points denoted $x_i \in \mathcal{M}$, in order that a fast Fourier transform could be performed. 
The frequencies for removal were manually selected so as to compromise between removal of splatter and avoidance of smoothing of the relevant variation in the dataset.
The resolution of the grid was taken to be 300$\times$300 (i.e. millimeter scale), which was fine enough to capture relevant variation whilst also enabling the computational analyses described in the sequel.

\subsubsection{Pre-Processed Dataset} \label{subsubsec: preproc dataset}

The result of pre-processing a panel dataset is a collection of triples
$$
(x_i,u_i^{(1)},u_i^{(2)})
$$
where $x_i \in \mathcal{M}_1$ represents a location in the manifold representation $\mathcal{M}_1$ of the panel, $u_i^{(1)} = u^{(1)}(x_i) \in \mathbb{R}$ represents the measured Euclidean distance of the upper surface of the panel from $x_i \in \mathcal{M}_1$ and $u_i^{(2)} = u^{(2)}(x_i) \in \mathbb{R}$ represents the measured Euclidean distance of the lower surface of the panel from $x_i \in \mathcal{M}_1$.
The subscript on $\mathcal{M}_1$ indicates that the training data are associated to this manifold; a test manifold $\mathcal{M}_2$ will be introduced in Section \ref{subsec: predict cylinder 1}.
The latent two-dimensional vector field $u$ describes the continuous geometry of the specific panel and it is the statistical nature of these vector fields which will be modelled.
In the sequel a collection of candidate models for $u$ are considered, each based on a Gaussian stochastic process.

The pre-processed data for a panel are collectively denoted $\mathcal{D}^{(j)}$, where $j \in \{1,\dots,6\}$ is the index of the panel.
The collection of all 6 of these panel datasets is denoted $\mathcal{D}_1$, the subscript indicating that these constitute the training dataset.

\subsection{Model Selection} \label{subsec: model select}

To perform formal model selection among the candidate models an information criterion was used, specifically the Akaike information criterion (AIC)
\begin{align}
\text{AIC}(M) = 2 \text{dim}(\hat{\theta}_M) - 2 \log p(\mathcal{D}_1 | \hat{\theta}_M , M) , \label{eq: AIC}
\end{align}
where $p(\mathcal{D}_1 | \theta_M , M)$ denotes the likelihood associated with the training dataset $\mathcal{D}_1$ under the SPDE model $M$, whose parameters are $\theta_M$, and $\hat{\theta}_M$ denotes a value of the parameter $\theta_M$ for which the likelihood is maximised. 
Small values of AIC are preferred.
The parameter vector $\theta_M$ can be decomposed into parameters that are identical for each panel dataset $\mathcal{D}^{(j)}$, denoted $\theta_M^\text{id}$, and those that are specific to a panel dataset $\mathcal{D}^{(j)}$, denoted $\theta_M^{(j)}$.
The latter are essentially random effects and consist of the Fourier coefficients $\gamma_\tau^{(\cdot)}$ and $\gamma_\eta^{(\cdot)}$ in the case where $M$ is a non-stationary model (Sections \ref{subsubsec: nonstat model} and \ref{subsubsec: ani nonstat}), otherwise $\theta_M^{(j)}$ is empty and no random effects are present.
The panel datasets $\mathcal{D}^{(j)}$ that together comprise the training dataset $\mathcal{D}_1$ are considered to be independent given $\theta_M$ and $M$, so that
\begin{align}
p(\mathcal{D}_1 | \theta_M, M) = \prod_{j=1}^6 p(\mathcal{D}^{(j)} | \theta_M^{\text{id}}, \theta_M^{(j)}, M) . \label{eq: likelihood}
\end{align}

The maximisation of \eqref{eq: likelihood} to obtain $\hat{\theta}_M$ requires a global optimisation routine and this will in general preclude the exact evaluation of the AIC for the statistical models considered.
Following standard practice, $\hat{\theta}_M$ was therefore set to be the output of a numerical optimisation procedure applied to maximisation of \eqref{eq: likelihood}. 
A drawback of the SPDE approach is that implementation of an effective numerical optimisation procedure is difficult due to the considerable computational cost involved in differentiating the likelihood.
To obtain numerical results, this work arrived, by trial-and-error, at a procedure that contains the following ingredients:

\vspace{5pt}
\noindent \textbf{Gradient ascent:} In the case where $M$ is a stationary model, so the $\theta_M^{(j)}$ are empty, the maximum likelihood parameters $\hat{\theta}_M^{\text{id}}$ were approximated using the natural gradient ascent method.
This extracts first order gradient information from \eqref{eq: likelihood} and uses the Fisher information matrix as a surrogate for second order gradient information, running a quasi-Newton method that returns a local maxima of the likelihood.
Full details are provided in Appendix A.4.

\vspace{5pt}
\noindent \textbf{Surrogate likelihood:} 
The use of gradient information constitutes a computational bottleneck, due to the large dense matrices involved, which motivates the use of a surrogate likelihood within the gradient ascent procedure whose Fisher information is more easily computed \citep{Tajbakhsh2014}.
For this paper a surrogate likelihood was constructed based on a subset of the dataset, reduced from the 300 $\times$ 300 grid to a 50 $\times$ 50 grid in the central portion of the panel. 
Natural gradient ascent was feasible for the surrogate likelihood.
It is hoped that minimisation of this surrogate likelihood leads to a suitable value $\hat{\theta}_M$ with respect to maximising the exact likelihood, and indeed the values that reported in the main text are evaluations of the exact likelihood at $\hat{\theta}_M$. 

\vspace{5pt}
\noindent \textbf{Two-stage procedure for random effects:} The case where $M$ is non-stationary is challenging for the natural gradient ascent method, even with the surrogate likelihood, due to the much larger dimension of the parameter vector when the random effects $\theta_M^{(j)}$ are included.
In this case a two-stage optimisation approach was used, wherein the common parameters $\theta_M^{\text{id}}$ were initially fixed equal to their values under the corresponding stationary model and then optimised over the random effects $\theta_M^{(j)}$ using natural gradient ascent, which can be executed in parallel. 

\vspace{5pt}
These approximation techniques combined to produce the most robust and reproducible parameter estimates that the authors were able to achieve, but it is acknowledged that they do not guarantee finding a local maximum of the likelihood and this analytic/numerical distinction can lead to pathologies that are discussed next. 
The computations that we report were performed in Matlab R2019b and required approximately one month of CPU time on a Windows 10 laptop with an Intel$^\text{\textregistered}$ Core\texttrademark i7-6600U CPU and 16GB RAM.
The number of iterations of natural gradient ascent required to achieve convergence ranged from 10 to 80 and was model-dependent.
Models with $\geq 100$ parameters required several days of computation, while the simpler models with $<10$ parameters typically required less than one day of computation in total.

\begin{table}[t]
\footnotesize
\centering
\begin{tabular}{|>{\centering\arraybackslash}p{1cm}>{\centering\arraybackslash}p{1cm}>{\centering\arraybackslash}p{1.7cm}cccc|} \hline
\multirow{2}{*}{\rotatebox{20}{Isotropic}} & \multirow{2}{*}{\rotatebox{20}{Stationary}} & \multirow{2}{*}{\rotatebox{20}{Noise}} & & \multicolumn{2}{c}{\textbf{Log-likelihood}} & \\
 &  &  & \textbf{\# Parameters} & surrogate & exact & \textbf{AIC} \\ \hline
\cmark & \cmark & white & 4 & 21,187 & 1,041,766 & -2,083,524 \\
\cmark & \cmark & smoother & 5 & 21,775 & 1,154,950 & -2,309,890 \\
\cmark & \cmark & sm. oscil. & 6 & 22,752 & 1,208,810 & -2,417,608 \\
\cmark & \xmark & white & 100 & 24,404 & 774,539  & -1,548,878  \\
\cmark & \xmark & smoother &101 & 25,900 & 1,023,561 & -2,046,920 \\
\cmark & \xmark & sm. oscil. & 102 & 29,118 & 1,042,737 & -2,085,270 \\
\xmark & \cmark & white & 6 & 22,005 & 1,100,358 & -2,200,704 \\  
\xmark & \cmark & smoother & 7 & 23,166 & 1,164,952 & -2,329,890 \\   
\xmark & \cmark & sm. oscil. & 8 & 25,151 & 1,233,936 & -2,467,856 \\   
\xmark & \xmark & white & 102 & 25,385 & 914,913 & -1,829,622 \\ 
\xmark & \xmark & smoother & 103 &  26,336 & 1,056,735 & -2,113,264 \\  
\xmark & \xmark & sm. oscil. & 104 & 28,150 & 1,111,449 & -2,222,690 \\ \hline  
\end{tabular}
\caption{Model selection based on the training dataset $\mathcal{D}_1$.
The first three columns indicate whether the differential operator $\mathcal{L}$ was isotropic and stationary and the driving noise process $\mathbf{Z}$ that was used.
The remaining columns indicate the number of free parameters in each model, the approximately maximised values of the surrogate and exact log-likelihood and the Akaike information criterion (AIC).
[The log-likelihood and AIC values are rounded to the nearest integer.]
}
\label{tab: AIC}
\end{table}

The twelve candidate models are presented in the first three columns of Table \ref{tab: AIC}, together with the dimension $\text{dim}(\theta_M)$ of their parameter vector, the (approximately) maximised value of the surrogate and exact log-likelihood and the AIC.
The main conclusions to be drawn are as follows:

\vspace{5pt}
\noindent \textbf{Isotropy:} In all cases the anisotropic models for $\mathcal{L}$ out-performed the corresponding isotropic models according to both the (approximately) maximised exact log-likelihood and the AIC. 
(The same conclusion held for 5 of 6 comparisons based on the surrogate likelihood.)
Since the anisotropic model includes the isotropic model as a special case, these higher values of the likelihood are to be expected, but the AIC conclusion is non-trivial.
(Inspection of the log-likelihood does, however, provide a useful validation of the optimisation procedure that was used.)

\vspace{5pt}
\noindent \textbf{Stationarity:} In all cases the non-stationary models for $\mathcal{L}$ out-performed the corresponding stationary models according to the (approximately) maximised surrogate log-likelihood, which is to be expected since these models are nested.
However, the conclusion was \emph{reversed} when the parameters $\hat{\theta}_M$, estimated using the surrogate likelihood, were used to evaluate the exact likelihood.
These results suggest that the two-stage numerical optimisation procedure for estimation of random effects was robust but that the surrogate likelihood was lossy in respect of information in the dataset that would be needed to properly constrain parameters corresponding to non-stationarity in the model.

\vspace{5pt}
\noindent \textbf{Noise model:} In all cases the smooth oscillatory noise models for $\mathbf{Z}$ led to substantially lower values of the surrogate and the exact log-likelihood compared to the white noise and smooth noise models.
This is to be expected since these models are nested, but the fact that this ordering was recovered in both the values of the surrogate and exact log-likelihood indicates that the numerical procedure worked well in respect of fitting parameters pertinent to the noise model.

\vspace{5pt}
\noindent \textbf{Best model:} 
The best-performing model, denoted $M^*$ in the sequel, was defined as the model that provided the smallest value of AIC, computed using the (approximately) maximised exact likelihood.
This was the anisotropic stationary model $\mathcal{L}$ with smooth oscillatory noise $\mathbf{Z}$, which had 8 parameters and achieved a substantially lower value of AIC compared to each of the 11 candidate models considered. 
Although the use of AIC in this context is somewhat arbitrary and other information criteria could also have been used, the substantially larger value of the log-likelihood for the best-performing model suggests that use of an alternative information criterion is unlikely to affect the definition of the best-performing model.

If instead the surrogate likelihood had been used to produce values for AIC, it would have declared that the best-performing model also includes non-stationary behaviour.
This suggests that the aforementioned ``best'' model is more accurately described as the ``best model that could be fitted'' and that, were the computational challenges associated with parameter estimation in the SPDE framework surmounted, a model that involves non-stationarity may have been selected. 
This suggests that additional models of further complexity may provide better descriptions of the data, at least according to AIC, but that fitting of model parameters presents a practical barrier to the set of candidate models that can be considered.
The remainder investigates whether predictions made using the simple model  $M^*$ may still be useful.

\section{Transfer Learning}

This section aims to assess the capacity of our selected statistical model $M^*$ to perform transfer learning, and proceeds as follows:
Section \ref{subsec: predict cylinder 1} assesses the performance of the statistical model in the context of a 3D-printed cylinder, which constitutes a held-out test dataset.
Then, Section \ref{subsec: predict cylinder 2} implements finite element simulations of compressive testing based on the fitted model and compares these against the results of a real compressive experiment.

\subsection{Predictive Performance Assessment} \label{subsec: predict cylinder 1}

To investigate the generalisation performance of the model $M^*$ identified in Section \ref{subsec: model select}, a test dataset, denoted $\mathcal{D}_2$ and displayed in Figure \ref{subfig: real cylinder}, was obtained as a laser scan of a 3D-printed cylinder of diameter $\approx$170mm and length $\approx$581mm. 
The notional thickness of the cylinder was 3.5mm, in agreement with the training dataset. 
The manifold representation of the panel is denoted $\mathcal{M}_1$ and the manifold representation of the cylinder is denoted $\mathcal{M}_2$.
The fitted model $M^*$ induces a bivariate Gaussian field on any suitably regular manifold, in particular $\mathcal{M}_2$.
A sample from this model is displayed in Figure \ref{subfig: cylinder sample}.
At a visual level, the ``extent'' of the variability is similar between Figure \ref{subfig: real cylinder} and Figure \ref{subfig: cylinder sample}, but the experimentally produced cylinder contains more prominent banding structure compared to the cylinder simulated from $M^*$.

The thickness of the cylinder wall is relevant to its mechanical performance when under load.
Recall that $M^*$ was not explicitly trained as a predictive model for thickness; rather, thickness is a derived quantity of interest. 
It is therefore interesting to investigate whether the distribution of material thickness, accouring to $M^*$ instantiated on $\mathcal{M}_2$, agrees with the experimentally obtained test dataset.
Results in Figure \ref{subfig: thickness} indicate that the fitted model agrees with the true distribution of material thickness in as far as the modal value and the right tail are well-approximated.
However, the true distribution is positively skewed and the model $M^*$, under which thickness is necessarily distributed as a Gaussian random field, is not capable of simultaneously approximating also the left tail.
From an engineering perspective, the model $M^*$ is conservative in the sense that it tends to under-predict wall thickness, in a similar manner to how factors of safety might be employed.

The predictive likelihood of a held-out test dataset is often used to assess the predictive performance of a statistical model.
However, there are two reasons why such assessment is unsuitable in the present context:
First, the pre-processing of the training dataset, described in Section \ref{subsec: preprocess}, was adapted to the manifold $\mathcal{M}_1$ of the training dataset and cannot be directly applied to the manifold $\mathcal{M}_2$ of the test dataset.
This precludes a fair comparison, since the fitted statistical model is not able to explain the high-frequency detail (such as weld splatter, visible in Figure \ref{subfig: real cylinder}) that are present in the test dataset $\mathcal{D}_2$ but were not included in the training dataset.
Second, recall that the purpose of the statistical model is limited to capturing aspects of geometric variation that are consequential in engineering. 
The predictive likelihood is not well-equipped to determine if the statistical model is suitable for use in the engineering context.

For a more detailed assessment of the statistical model and its predictive performance in the engineering context, finite element simulations of cylinders under load were performed, described next.

\begin{figure}[t!]
\centering
\begin{subfigure}[b]{0.3\textwidth}
\includegraphics[width = \textwidth,clip,trim = 0cm -4cm 0cm 0cm]{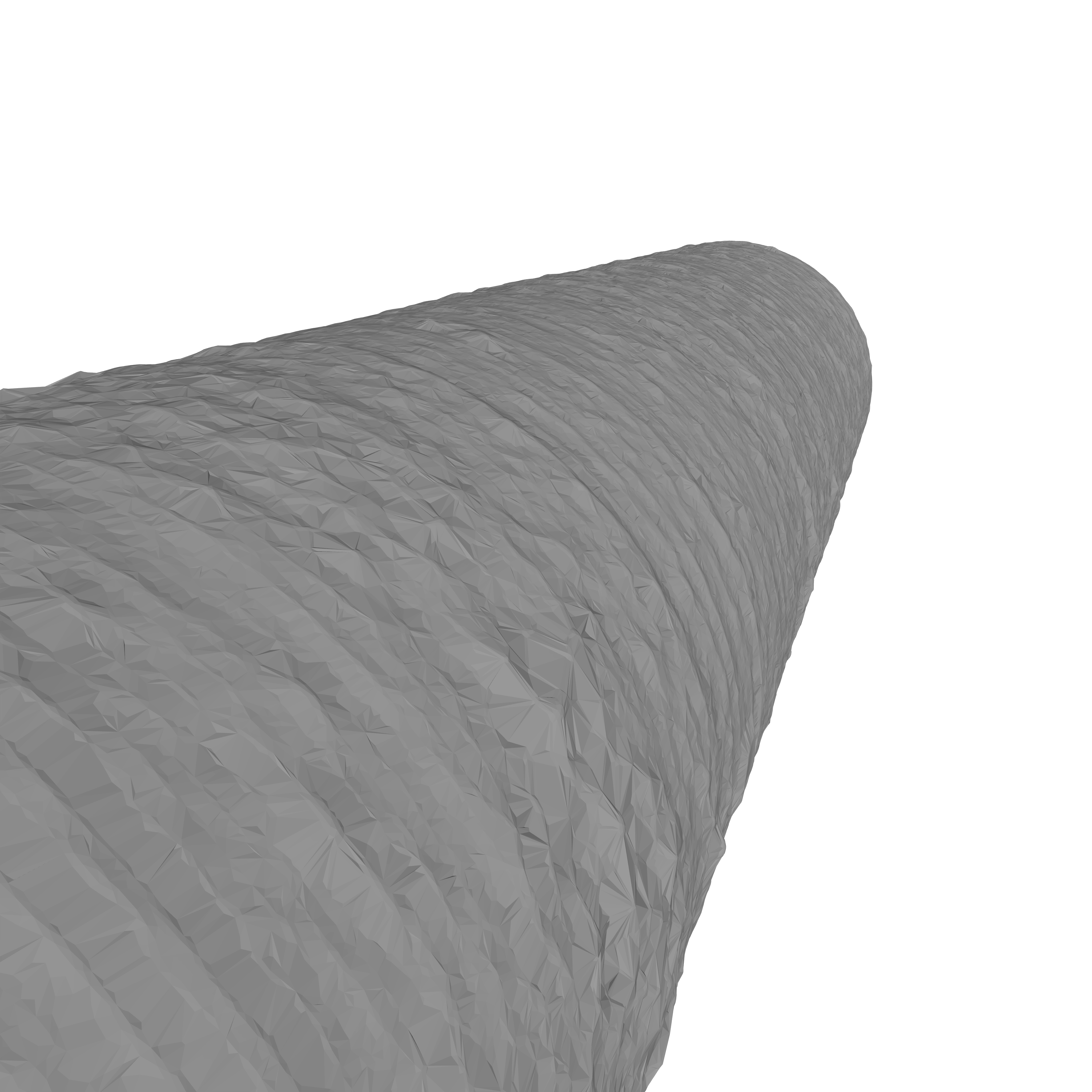}
\caption{}
\label{subfig: real cylinder}
\end{subfigure}
\begin{subfigure}[b]{0.3\textwidth}
\includegraphics[width = \textwidth,clip,trim = 0cm -4cm 0cm 0cm]{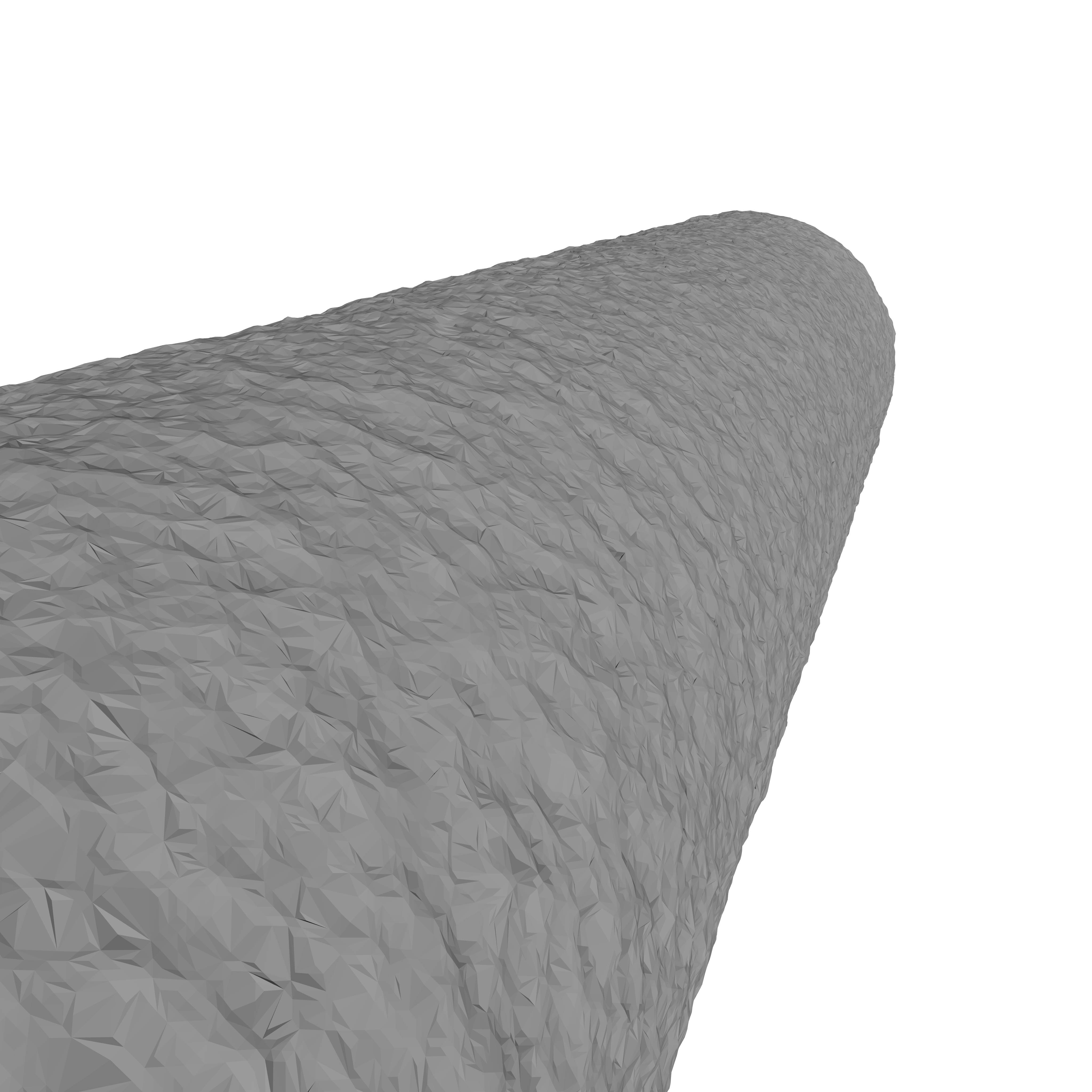}
\caption{}
\label{subfig: cylinder sample}
\end{subfigure}
\begin{subfigure}[b]{0.33\textwidth}
\includegraphics[width = \textwidth,clip,trim = 7cm 11.7cm 7cm 12cm]{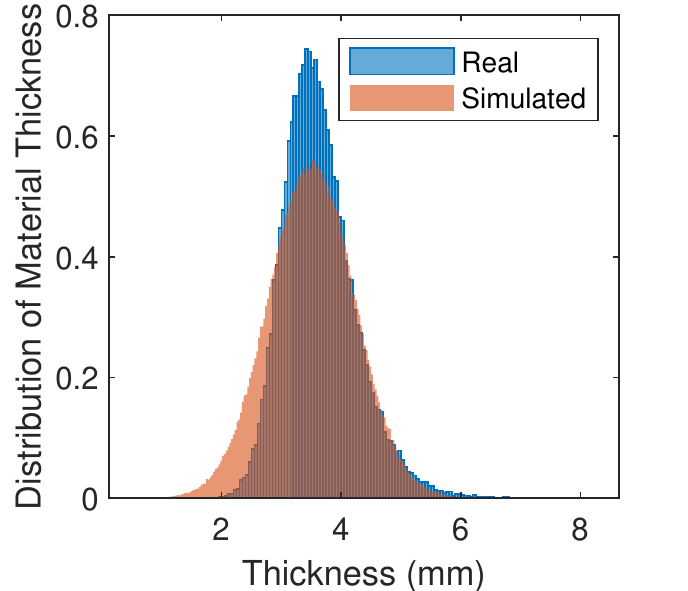}
\caption{}
\label{subfig: thickness}
\end{subfigure}
\caption{(a) Laser scan of a section from a real 3D-printed cylinder.
(b) Sample from the fitted model $M^*$ generated based on a cylindrical manifold.
(c) The distribution over the thickness of the wall of the real (blue, shaded) and the simulated (red, solid) 3D-printed cylinder.}
\label{fig: cylinder samples}
\end{figure}

\subsection{Application to Compressive Testing} \label{subsec: predict cylinder 2}

As discussed in Section \ref{subsec: candidates}, the aim of this work is to model those aspects of variability that are salient to the performance of components manufactured from 3D-printed steel.
To this end, predictions were generated for the outcome of a compressive test of a cylinder. 
This test, which was also experimentally performed, involved a cylinder of the same dimensions described in Section \ref{subsec: predict cylinder 1}, which was machined to ensure that its ends are parallel to fit the testing rig. 
The test rig consisted of a pair of end plates which compress the column ends axially. 
The cylinder was placed under load until a set amount of displacement has been reached. 
Both strain gauge and digital image correlation measurements of displacement under a given load were taken; further details may be found in \cite{Buchanan2018TESTINGOW}.

\begin{figure}[t!]
\centering
\begin{subfigure}[b]{0.19\textwidth}
\centering
\includegraphics[height = 0.3\textheight,clip,trim = 25cm 2cm 25cm 2cm]{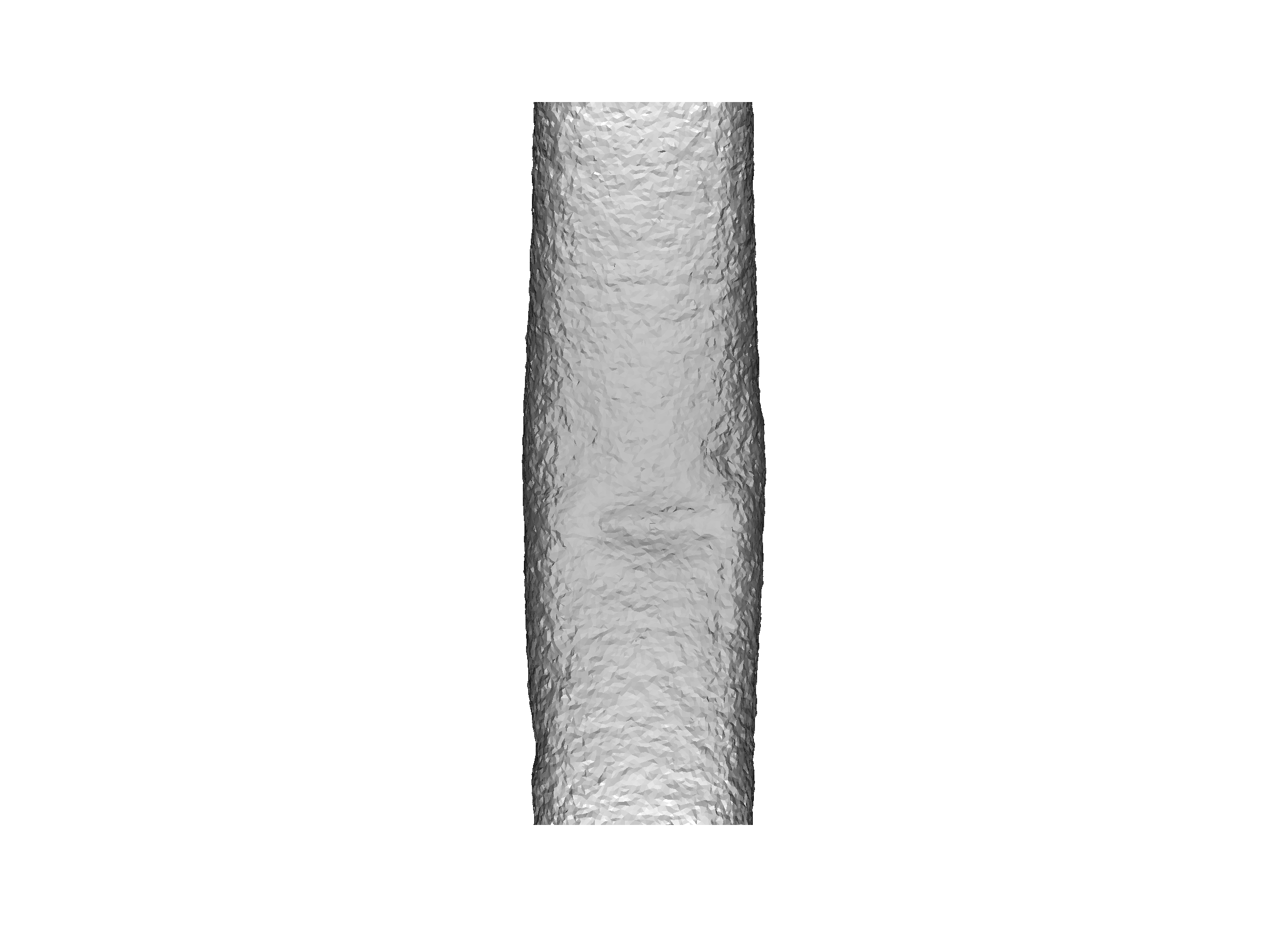}
\caption{}
\end{subfigure}
\begin{subfigure}[b]{0.19\textwidth}
\centering
\includegraphics[height = 0.3\textheight,clip,trim = 25cm 2cm 25cm 2cm]{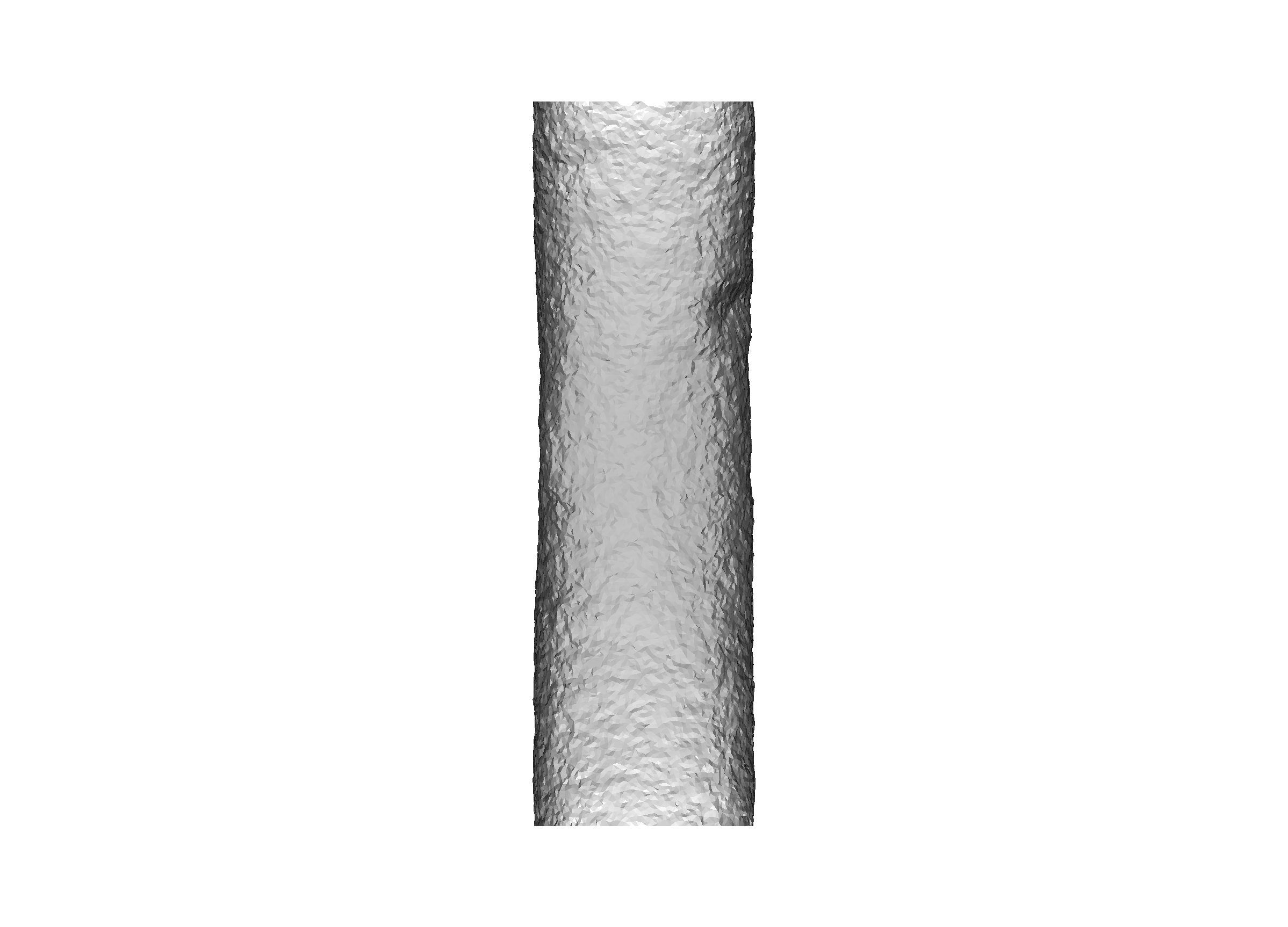}
\caption{}
\end{subfigure}
\begin{subfigure}[b]{0.19\textwidth}
\centering
\includegraphics[height = 0.3\textheight,clip,trim = 25cm 2cm 25cm 2cm]{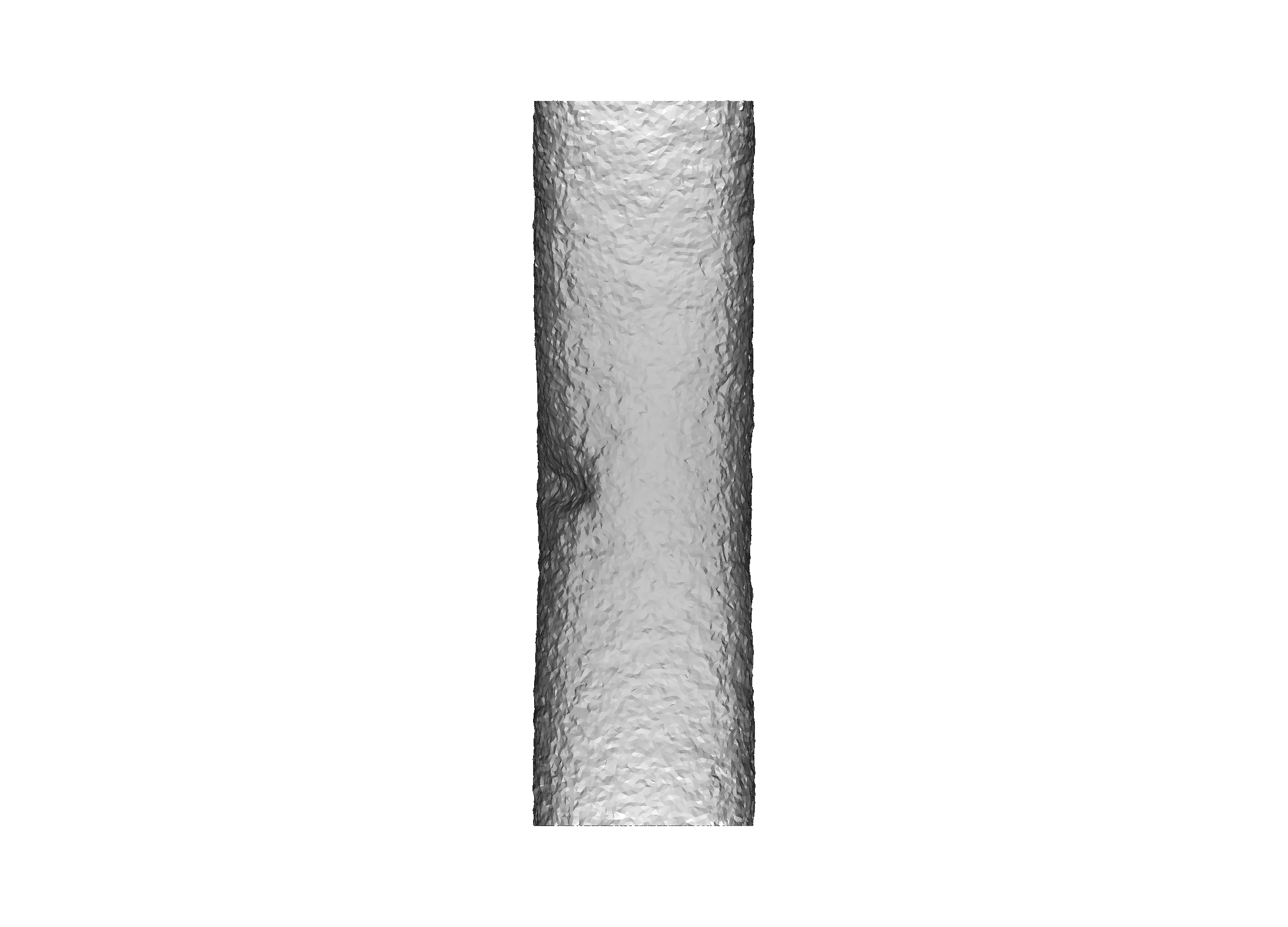}
\caption{}
\end{subfigure}
\begin{subfigure}[b]{0.19\textwidth}
\centering
\includegraphics[height = 0.3\textheight,clip,trim = 0cm 0cm 3.2cm 0cm]{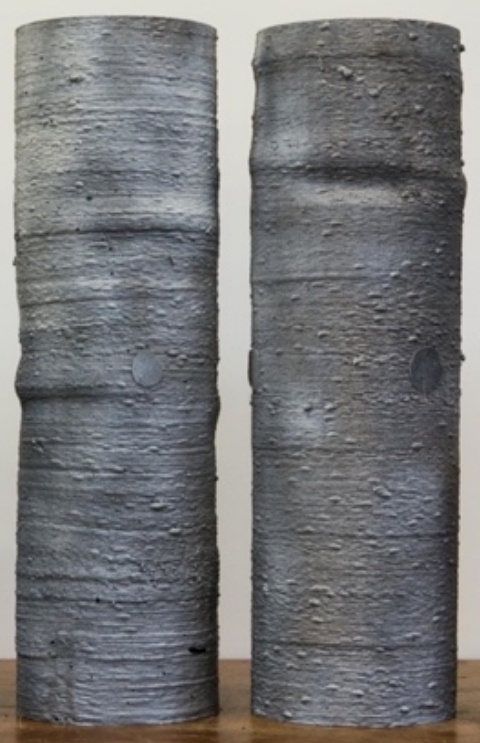}
\caption{}
\end{subfigure}
\begin{subfigure}[b]{0.19\textwidth}
\centering
\includegraphics[height = 0.3\textheight,clip,trim = 3cm 0cm 0cm 0cm]{figures/real_buckle}
\caption{}
\end{subfigure}
\caption{Predicting the outcome of a compressive test on a 3D-printed cylinder, using the fitted generative model $M^*$ for the geometry of the cylinder and a finite element simulation of the experiment.
(a-c) Samples from the generative model for the geometry of the cylinder after the simulated compressive test. 
(d-e) Buckled cylinders, experimentally produced. \citep[Reproduced with permission from][]{Buchanan2018TESTINGOW}.
}
\label{fig: FEM results}
\end{figure}

To produce a predictive distribution over possible outcomes of the test, three geometries were simulated for the cylinder using $M^*$ and finite element simulations of the compressive test were performed, using these geometries in the construction of the finite element model. 
To facilitate these simulations a simple homogeneous model for the mechanical properties of the material was used, whose values were informed by the coupon testing undertaken in \cite{Buchanan2018TESTINGOW}.
This is not intended as a realistic model for the mechanical properties of 3D-printed steel.
Indeed, as discussed in Section \ref{sec: intro}, it is expected that 3D-printed steel exhibits local variation in mechanical properties such as stiffness; the focus of this paper is surface metrology (only) and joint modelling of geometric and mechanical variation for 3D-printed steel will be addressed in future work.

Three typical end points from the finite element simulation are displayed in Figure \ref{fig: FEM results}, panels (a)-(c).
Notice that local variation in geometry leads to non-trivial global variation in the location of the buckling point.
Panel (d) in Figure \ref{fig: FEM results} displays the buckled states of two real 3D-printed steel cylinders, experimentally obtained.
At a visual level there is some qualitative agreement in the appearence of these failure modes.
The approach taken in this paper can be contrasted with approaches in the engineering literature that do not generalise across manifolds and do not report full probability distributions \citep{wagner2020probabilistic}.
However, an important caveat is that the precise quantitative relationship between load and displacement in these simulations is not yet meaningful due to the simplistic model used for material properties. 
Further research will be required to incorporate these aspects into a complete statistical model for 3D-printed steel.

\section{Discussion} \label{sec: conclusion}

This paper identified surface metrology as a promising new application area for recent advances in spatial statistics.
Though attention was focussed on 3D-printed stainless steel, the techniques and methods have the potential for application to a wide variety of problems in this engineering sub-field. 

Our approach sought to decouple the local statistical properties of the geometric variation from the global shape of the component, which offered the potential for transfer learning; i.e. conclusions drawn from one component to be carried across to a possibly different component.
This assumption is intuitively reasonable for components that do not involve high levels of curvature, but its appropriateness for general components remains to be tested.
To achieve this we exploited the SPDE framework.
However, these was a practical limitation on the complexity of SPDE that could be reliably fit, since gradients of the likelihood required large dense matrices to be computed. 
As a result, it is possible that models that are too simple but whose parameters can be successfully optimised may be prefered to more parsimonious models whose parameters cannot easily be optimised.
In other applications of surface metrology where generalisation across manifolds is not required, model fitting and model selection could be more easily achieved using the traditional covariance function framework \citep[e.g.][]{Rasmussen2006,Duvenaud2014,Bazavan2012,Oliva2016,Moeller2016}.

The principal mathematical assumption that made transfer learning possible in this work was the identification of Laplacians across different manifolds, corresponding to the different components that can in principle be manufactured.
This assumption (together with knowledge of the printing direction) enabled the meaning of a statistical model to be unambiguously interpreted across different manifolds, despite the fact that in general there are an infinitude of different geometries with which a given manifold can be equipped.
Gaussian process models arise naturally in this context, since in specific cases they can be conveniently expressed in terms of the Laplacian via the SPDE framework.
Alternative statistical models could also be considered:
For example, the eigenfunctions $\{E_k : k = 0,1,\dots\}$ of the negative Laplacian $-\Delta_{\mathcal{M}}$ form a basis for $L^2(\mathcal{M})$, and one can consider more general generative models for the random fields $u_1,u_2 \in L^2(\mathcal{M})$ based on linear combinations of the $E_k$ with non-Gaussian coefficients \citep[see e.g.][]{hosseini2017well,sullivan2017well}.
However, computational methods for non-Gaussian models remain relatively under-developed at present.

The scope of this investigation was limited to describing the geometric variation of 3D-printed steel and there are many interesting questions still to be addressed.
These include statistical modelling of the mechanical properties of the material, optimisation of the printing protocol to maximise the performance of (or reduce the variability in performance of) a given component, and adaptive monitoring and control of the manufacturing process to improve the quality of the printed component. 
These important problems will need to be collaboratively addressed; see \citet{Dodwell2021}.
The same printing protocol that we studied was recently used to construct an entire pedestrian bridge using 3D-printed steel \citep{Zastrow2020}.
To ensure the safety of the bridge for pedestrians, extensive engineering assessment and testing were required.
A long-term goal of this research is to enable such assessment to be partly automated {\it in silico} before the component is printed.
This would in turn reduce the design and testing costs associated with components manufactured with 3D-printed steel.

\paragraph*{Supplementary Materials}

The electronic supplement contains a detailed description of how computation was performed, together with the Matlab R2019b code that was used.

\paragraph*{Acknowledgements}
The authors wish to thank the Editor, Associate Editor, and two Reviewers for their valuable comments on the manuscript.

This research was supported by the Lloyd's Register Foundation programme on data-centric engineering and the Alan Turing Institute (under the EPSRC
grant EP/N510129/1).
WSK gratefully acknowledges support from the UK EPSRC (grant EP/K031066/1).
The authors are grateful to Craig Buchanan, Pinelopi Kyvelou and Leroy Gardner for providing the laser-scan datasets and to MX3D, from whom the material samples were provided.
In addition, the authors are grateful for discussions with Alessandro Barp, Tim Dodwell, Mark Girolami, Peter Gosling and Rob Scheichl.

} \fi

\newpage

\if1\supplement
{

\appendix
\setcounter{page}{1}
\section{Supplementary Material}

This appendix explains in detail how the computations described in the main text were performed.
The calculations are essentially identical to those described in \cite{Lindgren2011,Hu2013a,Hu2013}.
However, the authors believe it is valuable to include full details since previous work has left elements of these calculations implicit.
For instance, compared to the earlier work of \cite{Hu2013a,Hu2013}, this paper is explicit about how Neumann boundary conditions are employed. 

Appendix A.1 recalls how a general coupled system of SPDEs can be discretised using the classical finite element method. 
Then, in Appendix A.2 the model-specific aspects of the calculations are discussed.
Technical details regarding the choice of the finite elements are contained in Appendix A.3 and are essentially identical to Appendix A.2 in \cite{Lindgren2011}.
Lastly, in Appendix A.4 the natural gradient method is developed in the SPDE context and associated computational challenges are discussed.

\subsection{Discrete Representation of the Random Field} \label{ap: discrete rep of field}

This section discusses how a coupled system of SPDEs, as in Equation (2) of the main text, can be discretised using a finite element method.
Following \cite{Lindgren2011}, a finite-dimensional approximation
\begin{align}
\mathbf{u}_n(x) & = \sum_{l = 1}^2 \sum_{k=1}^n w_k^{(l)} \psi_k^{(l)}(x) \mathbf{e}_l , \label{eq: discrete approx to SPDE}
\end{align}
is used, where $\{\psi_k^{(l)}\}_{k=1}^n$ are basis functions to be specified and $\{\mathbf{e}_l\}_{l=1}^2$ is the standard basis for $\mathbb{R}^2$. 
The solution of Equation (2) is random and this randomness will be manifested in \eqref{eq: discrete approx to SPDE} through the weights $w_k^{(l)}$.
Thus in construction of an approximation $\mathbf{u}_n$, a distribution for these weights must be identified such that $\mathbf{u}_n$ is an accurate approximation of the random field.
To this end, let $\langle f , g \rangle := \langle f_1 , g_1 \rangle_\mathcal{M} + \langle f_2 , g_2 \rangle_\mathcal{M}$ for $f,g : \mathcal{M} \rightarrow \mathbb{R}^2$ and consider the Galerkin method, which selects the $w_k^{(l)}$ in order that the  equations
\begin{align}
\langle \phi_i^{(j)} \mathbf{e}_j , \mathcal{L} \mathbf{u}_n \rangle \stackrel{d}{=} \langle \phi_i^{(j)} \mathbf{e}_j , \mathbf{Z} \rangle , \qquad i \in \{ 1,\dots,n \}, j \in \{1,2\} \label{eq: Galerkin}
\end{align}
hold.
The notation $\stackrel{d}{=}$ indicates that the laws of the random variables are equal.
Here $\phi_i^{(j)}$ are test functions to be specified.
This work restricts attention to the choice $\phi_i^{(j)} = \psi_i^{(j)} = \psi_i$, meaning that the test functions and the basis functions coincide and that the same set of basis functions is used to describe both the upper and the lower surface of the material. 
The system \eqref{eq: Galerkin} fully determines the weights $w_k^{(l)}$ given a realisation of $\mathbf{Z}$, as clarified in the following proposition:

\begin{proposition} \label{prop: linear system}
The weights $w_k^{(l)}$ satisfy the linear system $\mathbf{K} \mathbf{w} \stackrel{d}{=} \mathbf{z}$, where $\mathbf{K}$, $\mathbf{w}$ and $\mathbf{z}$ have the block form
\begin{align*}
\mathbf{w} = \left[ \begin{array}{c} \mathbf{w}^{(1)} \\ \mathbf{w}^{(2)} \end{array} \right], \qquad \mathbf{z} = \left[ \begin{array}{c} \mathbf{z}^{(1)} \\ \mathbf{z}^{(2)} \end{array} \right], \qquad \mathbf{K} = \left[ \begin{array}{cc} \mathbf{K}^{(11)} & \mathbf{K}^{(12)} \\ \mathbf{K}^{(21)} & \mathbf{K}^{(22)} \end{array} \right] 
\end{align*}
and whose blocks are
\begin{align*}
\mathbf{w}^{(l)} = \left[ \begin{array}{c} w_1^{(l)} \\ \vdots \\ w_n^{(l)} \end{array} \right], \qquad \mathbf{z}^{(r)} = \left[ \begin{array}{c} \langle \phi_1^{(r)} , Z^{(r)} \rangle_\mathcal{M} \\ \vdots \\ \langle \phi_n^{(r)} , Z^{(r)} \rangle_\mathcal{M} \end{array} \right]
\end{align*}
and
\begin{align*}
\mathbf{K}^{(rs)} = \left[ \begin{array}{ccc} \langle \phi_1^{(r)} , \mathcal{L}^{(rs)} \psi_1^{(s)} \rangle_\mathcal{M} & \dots & \langle \phi_1^{(r)} , \mathcal{L}^{(rs)} \psi_n^{(s)} \rangle_\mathcal{M} \\ \vdots & & \vdots \\ \langle \phi_n^{(r)}, \mathcal{L}^{(rs)} \psi_1^{(s)} \rangle_\mathcal{M} & \dots & \langle \phi_n^{(r)} , \mathcal{L}^{(rs)} \psi_n^{(s)} \rangle_\mathcal{M} \end{array} \right] .
\end{align*}
\end{proposition}
\begin{proof}
Starting from the left hand side of \eqref{eq: Galerkin} one has that
\begin{align}
\langle \phi_i^{(j)} \mathbf{e}_j , \mathcal{L} \mathbf{u}_n \rangle & = \langle \phi_i^{(j)} , \mathcal{L}^{(j1)} (\mathbf{u}_n)_1 + \mathcal{L}^{(j2)}(\mathbf{u}_n)_2 \rangle_\mathcal{M} \nonumber \\
& = \left\langle \phi_i^{(j)} , \mathcal{L}^{(j1)} \left( \sum_{k=1}^n w_k^{(1)} \psi_k^{(1)} \right) + \mathcal{L}^{(j2)} \left( \sum_{k=1}^n w_k^{(2)} \psi_k^{(2)} \right) \right\rangle_\mathcal{M} \nonumber \\
& = \sum_{k=1}^n w_k^{(1)} \langle \phi_i^{(j)} , \mathcal{L}^{(j1)} \psi_k^{(1)} \rangle_\mathcal{M} + w_k^{(2)} \langle \phi_i^{(j)} , \mathcal{L}^{(j2)} \psi_k^{(2)} \rangle_\mathcal{M} \nonumber \\
& = [\mathbf{K}^{(j1)}]_{i, \cdot} \mathbf{w}^{(1)} + [\mathbf{K}^{(j2)}]_{i,\cdot} \mathbf{w}^{(2)} \nonumber \\
& = [\mathbf{K}^{(j1)} , \mathbf{K}^{(j2)}]_{i,\cdot} \mathbf{w}  \label{eq: start LHS}
\end{align}
and starting from the right hand side of \eqref{eq: Galerkin} one has that
\begin{align}
\langle \phi_i^{(j)} \mathbf{e}_j , \mathbf{Z} \rangle & = \langle \phi_i^{(j)} , Z^{(j)} \rangle_\mathcal{M} \nonumber \\
& = [\mathbf{z}^{(j)}]_i . \label{eq: RHS started}
\end{align}
Setting \eqref{eq: start LHS} equal to \eqref{eq: RHS started}, as enforced in \eqref{eq: Galerkin}, completes the proof.
\end{proof}

The following is immediate:

\begin{corollary} \label{cor: how to do it}
The precision matrix $\mathbf{Q} := \mathbb{C}(\mathbf{w})^{-1}$ for the weights is given by $\mathbf{Q} = \mathbf{K}^\top \mathbb{C}(\mathbf{z})^{-1} \mathbf{K}$.
\end{corollary}

The upshot of \Cref{cor: how to do it} is that, in order to implement the models discussed, one needs only to access to the matrix $\mathbf{K}$ and the precision matrix $\mathbb{C}(\mathbf{z})^{-1}$ of the random vector $\mathbf{z}$.
Formulae for each these are provided on a model-specific basis in Appendix A.2.
Note that $\mathbb{C}(\mathbf{z})$ will be block diagonal since $Z^{(1)}$ and $Z^{(2)}$ are independent.

Convergence of $\mathbf{u}_n$ to $u$ in the refined mesh limit $n \rightarrow 0$ was discussed in \cite{Lindgren2011}.
In addition to asymptotic convergence, at finite values of $n$ it is known that the discretisation just described suffers from variance inflation on the boundary of the meshed domain, compared to the actual random field.
This can be mitigated by meshing over a larger domain \citep[see Appendix A4 of][]{Lindgren2011}.
This was the approach taken when generating realisations of the cylinder in the main text; a longer cylinder was meshed and the middle portion was taken to represent the cylinder of interest.

\subsection{Calculations for Specific Model Components} \label{ap: specific models calc}

This appendix provides formulae for the matrices $\mathbf{K}$ and $\mathbb{C}(\mathbf{z})^{-1}$ (c.f. \Cref{cor: how to do it}). that are needed to discretise each of the twelve coupled systems of SPDEs considered.
The form of $\mathbf{K}$ will be determined by the choice of differential operator $\mathcal{L}$, and this is considered first, in Sections \ref{subsubsec: iso stat}-\ref{subsubsec: aniso nonstat}.
The distribution of $\mathbf{z}$ will be determined by the choice of noise process and this is considered second, in Sections \ref{subsubsec: white app}-\ref{subsubsec: so}. 

\subsubsection{Isotropic Stationary} \label{subsubsec: iso stat}

The isotropic stationary model has 
\begin{align*}
[\mathbf{K}^{(rs)}]_{ij} \defeq \langle \phi_i^{(r)} , \mathcal{L}^{(r,s)} \psi_j^{(s)} \rangle_\mathcal{M} & = \int \psi_i^{(r)} (\eta^{(rs)} - \Delta) \tau^{(rs)} \psi_j^{(s)} \mathrm{d} \mu_\mathcal{M} \\
& = \tau^{(rs)} \eta^{(rs)} \underbrace{ \int \psi_i^{(r)} \psi_j^{(s)} \mathrm{d} \mu_\mathcal{M} }_{\eqdef C_{ij}^{(rs)}} - \tau^{(rs)} \underbrace{ \int \psi_i^{(r)} \Delta \psi_j^{(s)} \mathrm{d} \mu_\mathcal{M} }_{\eqdef - G_{ij}^{(rs)}} 
\end{align*}
or, in matrix format,
\begin{align*}
\mathbf{K}^{(rs)} = \tau^{(rs)} (\eta^{(rs)} \mathbf{C}^{(rs)} + \mathbf{G}^{(rs)}) .
\end{align*}

\subsubsection{Anisotropic Stationary} \label{subsubsec: aniso stat}

The anisotropic stationary model has 
\begin{align*}
[\mathbf{K}^{(rs)}]_{ij} \defeq \langle \phi_i^{(r)} , \mathcal{L}^{(rs)} \psi_j^{(s)} \rangle_\mathcal{M} & = \tau^{(rs)} \eta^{(rs)} \underbrace{ \langle \psi_i^{(r)} , \psi_j^{(s)} \rangle_\mathcal{M} }_{C_{ij}^{(rs)}} - \tau^{(rs)} \underbrace{ \langle \psi_i^{(r)} , \nabla \cdot \mathbf{H}^{(rs)} \nabla  \psi_j^{(s)} \rangle_\mathcal{M} }_{\eqdef - \tilde{G}_{ij}^{(rs)} } 
\end{align*}
or, in matrix format,
\begin{align}
\mathbf{K}^{(rs)} & = \tau^{(rs)} ( \eta^{(rs)} \mathbf{C}^{(rs)} + \tilde{\mathbf{G}}^{(rs)} ) . \label{eq: K aniso stat}
\end{align}

\subsubsection{Isotropic Non-stationary} \label{subsubsec: iso nonstat}

For the isotropic non-stationary model, the approximation that $\eta^{(rs)}$ and $\tau^{(rs)}$ are locally constant over the support of any of the $\psi_i^{(r)}$ was used.
Thus, letting $x_i^{(r)}$ denote an aribitrary point in the support of $\psi_i^{(r)}$,
\begin{align*}
[\mathbf{K}^{(rs)}]_{ij} \defeq \langle \phi_i^{(r)} , \mathcal{L}^{(r,s)} \psi_j^{(s)} \rangle_\mathcal{M} & = \int \psi_i^{(r)} (\eta^{(rs)} - \Delta) (\tau^{(rs)} \psi_j^{(s)}) \mathrm{d} \mu_\mathcal{M} \\
& = \int \underbrace{ \eta^{(rs)} \tau^{(rs)} }_{\approx \text{ cst.}} \psi_i^{(r)} \psi_j^{(s)} \mathrm{d} \mu_\mathcal{M} - \int \psi_i^{(r)} \Delta ( \underbrace{ \tau^{(rs)} }_{\approx \text{cst.}} \psi_j^{(s)}) \mathrm{d} \mu_\mathcal{M} \\
& \approx \eta^{(rs)}(x_i^{(r)}) \tau^{(rs)}(x_i^{(r)}) \underbrace{ \int \psi_i^{(r)} \psi_j^{(s)} \mathrm{d} \mu_\mathcal{M} }_{C_{ij}^{(rs)}} - \tau^{(rs)}(x_i^{(r)}) \underbrace{ \int \psi_i^{(r)} \Delta \psi_j^{(s)} \mathrm{d} \mu_\mathcal{M} }_{-G_{ij}^{(rs)}} .
\end{align*}
Thus, in matrix format,
\begin{align*}
\mathbf{K}^{(rs)} & = \bm{\tau}^{(rs)} ( \bm{\eta}^{(rs)} \mathbf{C}^{(rs)} + \mathbf{G}^{(rs)} ) .
\end{align*}
Here $\bm{\tau}^{(rs)}$ is a diagonal matrix with $[\bm{\tau}^{(rs)}]_{i,i} \defeq \tau^{(rs)}(x_i^{(r)})$ and $\bm{\eta}^{(rs)}$ is a diagonal matrix with $[\bm{\eta}^{(rs)}]_{i,i} \defeq \eta^{(rs)}(x_i^{(r)})$.

\subsubsection{Anisotropic Non-stationary} \label{subsubsec: aniso nonstat}

For the anisotropic non-stationary model, the approximation that $\eta^{(rs)}$ and $\tau^{(rs)}$ are locally constant over the support of any of the $\psi_i^{(r)}$ was used. 
Thus, letting $x_i^{(r)}$ denote an aribitrary point in the support of $\psi_i^{(r)}$,
\begin{align*}
[\mathbf{K}^{(rs)}]_{ij} & \defeq \langle \phi_i^{(r)} , \mathcal{L}^{(rs)} \psi_j^{(s)} \rangle_\mathcal{M} \\
& = \langle \psi_i^{(r)} , \eta^{(rs)} \tau^{(rs)} \psi_j^{(s)} \rangle_\mathcal{M} - \langle \psi_i^{(r)} , \nabla \cdot \mathbf{H}^{(rs)} \nabla ( \tau^{(rs)} \psi_j^{(s)} ) \rangle_\mathcal{M}  \\
& = \int \underbrace{ \eta^{(rs)} \tau^{(rs)} }_{\approx \text{cst.}} \psi_i^{(r)} \psi_j^{(s)} \mathrm{d} \mu_\mathcal{M} - \int \psi_i^{(r)} \nabla \cdot \mathbf{H}^{(rs)} \nabla ( \underbrace{ \tau^{(rs)} }_{\approx \text{cst.}} \psi_j^{(s)} ) \mathrm{d} \mu_\mathcal{M} \\
& \approx \eta^{(rs)}(x_i^{(r)}) \tau^{(rs)}(x_i^{(r)}) \underbrace{ \int \psi_i^{(r)} \psi_j^{(s)} \mathrm{d} \mu_\mathcal{M} }_{ C_{ij}^{(rs)} } - \tau^{(rs)}(x_i^{(r)}) \underbrace{ \int \psi_i^{(r)} \nabla \cdot \mathbf{H}^{(rs)} \nabla \psi_j^{(s)} \mathrm{d} \mu_\mathcal{M} }_{ - \tilde{G}_{ij}^{(rs)} } .
\end{align*}
Thus, in matrix format,
\begin{align}
\mathbf{K}^{(rs)} & = \bm{\tau}^{(rs)} ( \bm{\eta}^{(rs)} \mathbf{C}^{(rs)} + \tilde{\mathbf{G}}^{(rs)} ) . \label{eq: most general K}
\end{align}

\subsubsection{White Noise} \label{subsubsec: white app}

For the white noise model, it follows immediately from the definition that, since $\phi_i^{(j)} = \psi_i^{(j)}$, 
\begin{align*}
\mathbb{C}( \langle \phi_i^{(r)} , W^{(r)} \rangle_\mathcal{M} , \langle \phi_j^{(s)} , W^{(s)} \rangle_\mathcal{M} ) & = \delta_{rs} \langle \phi_i^{(r)} , \phi_j^{(s)} \rangle_\mathcal{M} \\
& = \delta_{rs} \langle \psi_i^{(r)} , \psi_j^{(s)} \rangle_\mathcal{M} \\
& = \delta_{rs} C_{ij}^{(rs)} .
\end{align*}
Thus, in matrix format, $\mathbb{C}(\mathbf{z}^{(r)}) = \mathbf{C}^{(rr)}$ for this noise model.

\subsubsection{Smoother Noise}

For the smoother noise model, observe that the system of SPDEs that governs the noise process, in Equation (7) of the main text, has the same generic form as the original system of SPDEs of interest. 
The same discretisation techniques described in \Cref{ap: discrete rep of field} can therefore be applied and an argument essentially identical to \Cref{prop: linear system} establishes that 
\begin{align*}
\mathbb{C}(\mathbf{z}^{(r)})^{-1} = ( \mathbf{K}^{(rr)} )^\top (\mathbf{C}^{(rr)})^{-1} \mathbf{K}^{(rr)}
\end{align*}
for this noise model.
Here $\mathbf{K}^{(r,s)}$ is as in \eqref{eq: K aniso stat}.

\subsubsection{Smooth Oscillatory Noise} \label{subsubsec: so}

For the smooth oscillatory noise model it can be shown that $Z^{(1)}$ and $Z^{(2)}$ are independent and each corresponds to an oscillatory Gaussian random field with
\begin{align*}
\mathbb{C}(\mathbf{z}^{(r)})^{-1} & = \eta^2 \mathbf{C}^{(rr)} + 2 \eta \cos(\pi \theta) \tilde{\mathbf{G}}^{(rr)} + \tilde{\mathbf{G}}^{(rr)} [\mathbf{C}^{(rr)}]^{-1} \tilde{\mathbf{G}}^{(rr)} .
\end{align*}
See Section 3.3 of \cite{Lindgren2011} for further detail.

\subsection{Calculations for Finite Elements} \label{app: fin elem calcs}

The purpose of this technical appendix is to provide explicit formulae for the matrices $\mathbf{C}$, $\mathbf{G}$ and $\tilde{\mathbf{G}}$ appearing in \Cref{ap: specific models calc}.
This is included merely for completeness and almost exactly follows \cite{Lindgren2011}.

First note that, for the matrix $\mathbf{G}$ appearing in isotropic models, one can use Green's first identity to obtain 
\begin{align*}
G_{ij}^{(rs)} & \defeq - \int \psi_i^{(r)} \Delta \psi_j^{(r)} \mathrm{d} \mu_\mathcal{M} \\
& = \int \nabla \psi_i^{(r)} \cdot \nabla \psi_j^{(s)} \mathrm{d} \mu_\mathcal{M} - \underbrace{ \oint \psi_i^{(r)} \nabla \psi_j^{(s)} \cdot \mathrm{n} \mathrm{d}\mu_{\partial \mathcal{M}} }_{=0 \text{ (Neumann b.c.)}}  \\
& = \langle \nabla \psi_i^{(r)} , \nabla \psi_j^{(s)} \rangle_\mathcal{M} .
\end{align*}
For the more general form $\tilde{\mathbf{G}}(\mathbf{H})$ appearing in anisotropic models, based on the weighted Laplacian $\nabla \cdot \mathbf{H} \nabla$, one can apply the divergence theorem to the vector field $\mathbf{F} = \psi_i^{(r)} \mathbf{H} \nabla \psi_j^{(s)}$ to obtain that 
\begin{align*}
\tilde{G}_{ij}^{(rs)}(\mathbf{H}) & \defeq - \int \psi_i^{(r)} \nabla \cdot \mathbf{H} \nabla \psi_j^{(r)} \mathrm{d} \mu_\mathcal{M} \\
& = \int \nabla \psi_i^{(r)} \cdot \mathbf{H} \nabla \psi_j^{(s)} \mathrm{d} \mu_\mathcal{M} - \underbrace{ \oint \psi_i^{(r)} \mathbf{H} \nabla \psi_j^{(s)} \cdot \mathrm{n} \mathrm{d}\mu_{\partial \mathcal{M}} }_{=0 \text{ (Neumann b.c.)}} \\
& = \langle \mathbf{H}^{1/2} \nabla \psi_i^{(r)} , \mathbf{H}^{1/2} \nabla \psi_j^{(s)} \rangle_\mathcal{M} .
\end{align*}
Thus computation of $\mathbf{C}$, $\mathbf{G}$ and $\tilde{\mathbf{G}}$ amounts to computing particular inner products over the manifold $\mathcal{M}$.

The finite element construction renders such computation straightforward by enforcing sparsity in these matrices, that in turn leads to sparsity of the matrix $\mathbf{K}$.
To construct finite elements the locations $x_i$ at which the data are defined (see main text) are used to produce a triangulation $\mathcal{M}^{(n)}$ of size $n$ that forms an approximation to $\mathcal{M}$ in $\mathbb{R}^3$. 
Let the vertices of the triangulation $\mathcal{M}^{(n)}$ be denoted $\mathbf{v}_1,\dots, \mathbf{v}_n$.
The basis functions $\psi_k$ were then taken to be the canonical piecewise linear finite elements associated with the triangulation $\mathcal{M}_n$, such that $\psi_k$ is centred at $\mathbf{v}_k \in \mathbb{R}^3$.
Consider a triangle $T = (\mathbf{v}_i,\mathbf{v}_j,\mathbf{v}_k)$ and denote the edges $\mathbf{e}_i = \mathbf{v}_k - \mathbf{v}_j$, $\mathbf{e}_j = \mathbf{v}_i - \mathbf{v}_k$, $\mathbf{e}_k = \mathbf{v}_j - \mathbf{v}_i$.
Let $\langle \cdot , \cdot \rangle$ denote the standard inner product on $\mathbb{R}^3$ and $\|\cdot\|$ its associated norm.
Then the area of $T$ is $|T| = \| \mathbf{e}_i \times \mathbf{e}_j \|/2$.
As explained in Appendix A2 of \cite{Lindgren2011}, the matrices $\mathbf{C}$, $\mathbf{G}$ and $\tilde{\mathbf{G}}$ are obtained from summing the contributions from each triangle; the contribution from triangle $T$ is
\begin{align*}
C_{\{i,j,k\} \times \{i,j,k\}} & \leftarrow C_{\{i,j,k\} \times \{i,j,k\}} + \frac{|T|}{12} \left[ \begin{array}{ccc} 2 & 1 & 1 \\ 1 & 2 & 1 \\ 1 & 1 & 2 \end{array} \right] \\
G_{\{i,j,k\} \times \{i,j,k\}} & \leftarrow G_{\{i,j,k\} \times \{i,j,k\}} + \frac{1}{4|T|} \left[ \begin{array}{ccc} \mathbf{e}_i & \mathbf{e}_j & \mathbf{e}_k \end{array} \right]^\top \left[ \begin{array}{ccc} \mathbf{e}_i & \mathbf{e}_j & \mathbf{e}_k \end{array} \right] \\
\tilde{G}_{\{i,j,k\} \times \{i,j,k\}} & \leftarrow G_{\{i,j,k\} \times \{i,j,k\}} + \frac{1}{4|T|} \left[ \begin{array}{ccc} \mathbf{e}_i & \mathbf{e}_j & \mathbf{e}_k \end{array} \right]^\top \text{adj}(\mathbf{H}) \left[ \begin{array}{ccc} \mathbf{e}_i & \mathbf{e}_j & \mathbf{e}_k \end{array} \right]
\end{align*}
where $\text{adj}(\mathbf{H}) = \text{det}(\mathbf{H}) \mathbf{H}^{-1}$ is the adjugate matrix of $\mathbf{H}$.

In general $\mathbf{C}^{-1}$ and hence $\mathbf{Q}$ are dense matrices, which presents a computational barrier to Cholesky factorisation of $\mathbf{Q}$.
Following \cite{Lindgren2011} the matrix $\mathbf{C}$ was replaced with the diagonal matrix $\tilde{\mathbf{C}}$ such that $\tilde{C}_{i,i} = \langle \psi_i , 1 \rangle_\mathcal{M}$.
Computationally, one has 
\begin{align*}
\tilde{C}_{\{(i,i), (j,j), (k,k)\}} & \leftarrow \tilde{C}_{\{(i,i), (j,j), (k,k)\}} + \frac{|T|}{3} \left[ \begin{array}{ccc} 1 & 1 & 1 \end{array} \right]  .
\end{align*}
It was argued in \cite{Lindgren2011} that the approximation of $\mathbf{C}$ by $\tilde{\mathbf{C}}$ still ensures convergence of the approximation of $\mathbf{u}_n$ to $u$ in the limit $n \rightarrow \infty$ as the mesh is refined.

\subsection{Natural Gradient Ascent} \label{subsec: natural gradient}

This appendix described the natural gradient ascent method used to (approximately) maximise the likelihood.

Let $\mathcal{D}$ be a dataset as described in the main text and let $\mathbf{y} \in \mathbb{R}^N$ be a vectorised representation of $\mathcal{D}$ following the block matrix convention of \Cref{prop: linear system}, so that $N = 2n$.
Suppose that the parameters $\theta_M$ of a model $M$ have been vectorised so that $\theta_M$ is represented by a vector $\bm{\theta} \in \mathbb{R}^p$.
The log-likelihood of a model $M$, with precision matrix $\mathbf{Q}(\bm{\theta})$ parametrised by $\bm{\theta}$, is
\begin{align*}
\ell(\bm{\theta}) \defeq \log p(\mathcal{D} | \theta_M , M ) & = - \frac{N}{2} \log(2\pi) + \frac{1}{2} \log \det \mathbf{Q}(\bm{\theta}) - \frac{1}{2} \mathbf{y}^\top \mathbf{Q}(\bm{\theta}) \mathbf{y} .
\end{align*}
The natural gradient method, due to \cite{Amari1998}, takes an initial guess $\bm{\theta}_0$ for $\bm{\theta}$ and then iteratively updates this guess according to
\begin{align*}
\bm{\theta}_{n+1} & = \bm{\theta}_n + \mathbf{I}(\bm{\theta}_n)^{-1} \partial_{\bm{\theta}} \ell(\bm{\theta}_n)
\end{align*}
Here $\partial_{\bm{\theta}} \ell$ is the gradient of the log-likelihood with components
\begin{align}
\partial_{\theta_i} \ell(\bm{\theta}) & = \frac{1}{2} \tr \left( \mathbf{Q}(\bm{\theta})^{-1} \frac{\partial \mathbf{Q}}{\partial \theta_i}(\bm{\theta}) \right) - \frac{1}{2} \mathbf{y}^\top \frac{\partial \mathbf{Q}}{\partial \theta_i}(\bm{\theta}) \mathbf{y} \label{eq: llgrads}
\end{align}
and $\mathbf{I}(\bm{\theta})$ is the Fisher information matrix with components
\begin{align*}
I_{ij}(\bm{\theta}) & = \frac{1}{2} \tr \left( \mathbf{Q}(\bm{\theta})^{-1} \frac{\partial \mathbf{Q}}{\partial \theta_i}(\bm{\theta}) \mathbf{Q}(\bm{\theta})^{-1} \frac{\partial \mathbf{Q}}{\partial \theta_j}(\bm{\theta}) \right) .
\end{align*}
The required derivatives $\partial_{\theta_i} \mathbf{Q}$ can be computed using matrix calculus to obtain the identities (letting $\partial = \partial_{\theta_i}$ denote differentiation with respect to a specific element $\theta_i$ of $\bm{\theta}$)
\begin{align*}
\partial \mathbf{Q} & = (\partial \mathbf{K})^\top \mathbb{C}(\mathbf{z})^{-1} \mathbf{K} - \mathbf{K}^\top \mathbb{C}(\mathbf{z})^{-1} (\partial \mathbb{C}(\mathbf{z})) \mathbb{C}(\mathbf{z})^{-1} \mathbf{K} + \mathbf{K}^\top \mathbb{C}(\mathbf{z})^{-1} (\partial \mathbf{K}) \\
\partial \mathbf{K}^{(rs)} & = (\partial \bm{\tau}^{(rs)}) (\bm{\eta}^{(rs)} \mathbf{C}^{(rs)} + \tilde{\mathbf{G}}^{(rs)}) + \bm{\tau}^{(rs)} ((\partial \bm{\eta}^{(rs)}) \mathbf{C}^{(rs)} + \partial \tilde{\mathbf{G}}^{(rs)}) 
\end{align*}
Here $\mathbf{K}$ is the most general form of the matrices variously called $\mathbf{K}$, presented in \eqref{eq: most general K}.
The natural gradient method can be viewed as a quasi-Newton method that, if it converges, converges to a local maximum of the likelihood.

An anonymous Reviewer helpfully pointed out that common sparsity structure in $\mathbf{Q}$ and $\partial_{\theta_i} \mathbf{Q}$ can be exploited by using Takahashi recursions to circumvent computation of the full dense matrix $\mathbf{Q}^{-1} \partial_{\theta_i} \mathbf{Q}$ in \eqref{eq: llgrads}, since only the trace is required \citep{zammit2018sparse}.
However, this sparsity exploit is not immediately applicable to the Fisher information matrix.
Indeed, the natural gradient ascent method requires a product of two matrices of the form $\mathbf{Q}^{-1} \partial_{\theta_i} \mathbf{Q}$, which unfortunately will be dense in general. 
One solution is simply to use the usual Euclidean gradient instead, which corresponds to performing standard gradient descent.
However, this carries the substantial disadvantage that a step-size parameter must be introduced and manually tuned.
An alternative solution is to develop an approximation strategy to circumvent the dense matrix computations; this has been considered in \cite{Tajbakhsh2014}.
In the present context, natural gradient ascent is performed only on the manifold $\mathcal{M}_1$ of the training dataset, whose notional flat geometry is easily meshed, and dense matrix computation is not required for subsequent predictions to be produced on the potentially large and complex test manifold $\mathcal{M}_2$.
For this reason one may be willing to perform computation with dense matrices when fitting the training dataset. 

} \fi

\end{document}